\documentclass[11pt]{article}
\usepackage{graphicx}

\usepackage{amsmath}
\usepackage{amsxtra}
\usepackage{amstext}
\usepackage{amssymb}
\usepackage{latexsym}
\usepackage{dsfont}

\begin{document}

\newcommand{\be}{\begin{equation}}
\newcommand{\ee}{\end{equation}}

\newcommand{\bn}{\begin{eqnarray}}
\newcommand{\en}{\end{eqnarray}}

\title{Fermion masses and neutrino mixing in an $U(1)_H$ flavor symmetry
model with hierarchical radiative generation for light charged
fermion masses}

\author{Albino Hern\'andez-Galeana\footnote{e-mail: albino@esfm.ipn.mx}\\
Departamento de F\'{\i}sica de la Escuela Superior de F\'{\i}sica
y Matem\'aticas\\
 Instituto Polit\'ecnico Nacional. U. P. "Adolfo
L\'opez Mateos". C. P. 07738.\\
 M\'exico, D. F.}

\maketitle

\vspace{5mm}

\begin{abstract}
I report the analysis performed on fermion masses and mixing,
including neutrino mixing, within the context of a model with
hierarchical radiative mass generation mechanism for light charged
fermions, mediated by exotic scalar particles at one and two
loops, respectively, meanwhile the neutrinos get Majorana mass
terms at tree level through the Yukawa couplings with two
$SU(2)_L$ Higgs triplets. All the resulting mass matrices in the
model, for the u, d, and e fermion charged sectors, the neutrinos
and the exotic scalar particles, are diagonalized in exact
analytical form. Quantitative analysis shows that this model is
successful to accommodate the hierarchical spectrum of masses and
mixing in the quark sector as well as the charged lepton masses.
The lepton mixing matrix, $V_{PMNS}$, is written completely in
terms of the neutrino masses $m_1, m_2$, and $m_3$. Large lepton
mixing for $\theta_{12}$ and $\theta_{23}$ is predicted in the
range of values $0.7 \lesssim \sin^22\theta_{12}\lesssim 0.7772$
and $0.87 \lesssim \sin^22\theta_{23} \lesssim 0.9023$ by using
$0.033 \lesssim s_{13}^2 \lesssim 0.04$. These values for lepton
mixing are consistent with $3 \sigma$ allowed ranges provided by
recent global analysis of neutrino data oscillation. From $\Delta
m_{\text{sol}}^2$ bounds, neutrino masses are predicted in the
range of values $m_1 \thickapprox (1.706 - 2.494) \times 10^{- 3}
eV$, $m_2 \thickapprox (6.675 - 12.56) \times 10^{- 3} eV$, and
$m_3 \thickapprox (1.215 - 2.188) \times 10^{- 2} eV$,
respectively. The above allowed lepton mixing leads to the
quak-lepton complementary relations $\theta_{12}^{CKM} +
\theta_{12}^{PMNS} \thickapprox 41.543^{\circ} - 44.066^{\circ}$
and $\theta_{23}^{CKM} + \theta_{23}^{PMNS} \thickapprox
36.835^{\circ} - 38.295^{\circ}$. The new exotic scalar particles
induce flavor changing neutral currents and contribute to lepton
flavor violating processes such as $E \rightarrow e_1 e_2 e_3$, to
radiative rare decays, $\tau \rightarrow \mu \gamma , \tau
\rightarrow e \gamma , \mu \rightarrow e \gamma$, as well as to
the anomalous magnetic moments of fermions. I give general
analytical expressions for the branching ratios of these rare
decays and for the anomalous magnetic moments for charged leptons.
\end{abstract}

\vspace{5mm}

Keywords: Neutrino mixing, Fermion masses and mixing, Flavor
symmetry.

\vspace{5mm} PACS: 14.60.Pq, 12.15.Ff, 12.60.-i

\section{Introduction}

The observed hierarchical spectrum of masses and mixing angles in
the quark sector still remains as one of the most important
challenges in particle physics. A possible solution to explain
this hierarchical spectrum is that light fermion masses arise
through radiative corrections \cite{mrad}, while the masses for
top quark, bottom quark, and tau lepton are generated either at
the tree level like in Ref.\cite{u1model}, or by the
implementation of seesaw-type mechanisms as was proposed by the
author in a model with a $SU(3)$ horizontal symmetry in Ref.
\cite{su3models}. Recently, it has also been possible to observe
the phenomenon of flavor mixing in the leptonic sector through the
confirmation of the phenomenon of "neutrino oscillation" in
experiments of neutrinos coming from atmospheric \cite{nuatm},
solar \cite{nusolar}, reactors \cite{nureac} and accelerators
\cite{nuacc}, and as a consequence the
Pontecorvo-Maki-Nakagawa-Sakata (PMNS) lepton mixing matrix,
$V_{PMNS}$, has been determined. For three flavor neutrinos $\nu_e
\;,\nu_\mu$, and $\nu_\tau$, which may evolve into the three
massive neutrinos $\nu_1\;,\nu_2\;,\nu_3$, the experiments of
neutrino oscillations are interpreted in terms of three mixing
angles denoted by $\theta_{12}$ for $\nu_e - \nu_\mu$,
$\theta_{23}$ for $\nu_\mu - \nu_\tau$, and $\theta_{13}$ for
$\nu_e - \nu_\tau$. Recent analysis and fit of neutrino
oscillation data give \cite{mohapatra}, at the 3$\sigma$ level,
the allowed ranges of values

\vspace{3mm}

\be \begin{array}{lcl} |{\Delta m}_{23}^2|=(1.4 - 3.3) \times
{10}^{-3}(eV)^2 & , & {\Delta m}_{12}^2 = (7.1 - 8.9) \times {10}^{-5}(eV)^2 \;, \\
\sin^22\theta_{23}=0.87 - 1.0  & , & \sin^22\theta_{12}= 0.70 - 0.94          \;,  \\
\sin^2\theta_{13} \leqslant 0.051 \;,  & & \end{array} \ee

\vspace{3mm}

\noindent where ${\Delta m}_{12}^2$ and ${\Delta m}_{23}^2$ are
the solar and atmospheric mass differences, respectively.

\vspace{5mm} In this article I address the problem of fermion
masses and mixing angles, including neutrino mixing, within the
context of the model introduced in Ref.[2]. Section 2 briefly
reviews the main features of the model with an $U(1)_H$ flavor
symmetry. Next, in Sec. 3 I discuss the masses and mixing for
charged leptons and quarks, at one and two loops, and using the
strong hierarchy of masses, approximate mixing matrices for the u,
d, and e charged sectors are provided. Section 4 is devoted to
finding the upper bounds for mixing angles of charged leptons. In
Sec. 5 I analyze neutrino mixing, and the $V_{PMNS}$ lepton mixing
matrix is written in terms of neutrino masses, providing numerical
results for neutrino mixing. In Sec. 6 I perform a quantitative
analysis of quarks masses and mixing, including numerical values
for the $V_{CKM}$. In Sec. 7 I give general expressions for the
flavor changing neutral currents (FCNCs), rare decays and
anomalous magnetic moments for charged leptons that are induced by
the exotic scalar particles. Section 8 contents my conclusions. In
the Appendix I have introduced a method to diagonalize in close
analytical form a generic 3x3 real and symmetric mass matrix, and
then I have extended this method to diagonalize the 4x4 real
symmetric exotic scalar mass matrix.

\section{Model with $U(1)_H$ flavor symmetry}

The gauge symmetry of the model is defined as $U(1)_H \otimes
SU(3)_C \otimes SU(2)_L \otimes U(1)_Y $. The fermionic content of
the model is the same as in the "standard model" (SM), and their
charges under the flavor symmetry $U(1)_H$ are arranged as to
cancel anomalies without the introduction of exotic fermions. The
fermions are classified, as in the SM, in five sectors $f = q, u,
d, l$, and $e$, where $q$ and $l$ are the $SU(2)_L$ quark and
lepton doublets, respectively, and $u, d$, and $e$ are the
singlets, in an obvious notation.

\vspace{3mm}
 The cancelation of anomalies in a simple way that
simultaneously guarantees that only the third gene\-ration of the
charged fermions acquire masses at tree level is given by
\cite{u1model}

\bn H(f)=0,\pm \delta_f &,&
\delta_q^2-2\delta_u^2+\delta_d^2=\delta_l^2-\delta_e^2 \:,\en

\noindent with the constraints

\begin{equation}
\delta_\textit{l}=\delta_q=\Delta\neq\delta_u=\delta_d=\delta_e=\delta
\end{equation}

\vspace{3mm}
 The assignment of flavor charges to the fermions is
then as given in Table 1. The $G_{SM} \equiv SU(3)_C \otimes SU(2)_L
\otimes U(1)_Y$ quantum numbers of fermions are the same as in the
SM.

\begin{table}
\begin{center}
\begin{tabular}{|c|c|c|c|}\hline
Sector&Family 1&Family 2&Family 3\\ \hline\hline
$q$&$\Delta$&$-\Delta$&0\\
$u$&$\delta$&$-\delta$&0\\
$d$&$\delta$&$-\delta$&0\\
$\textit{l}$&$\Delta$&$-\Delta$&0\\
$e$&$\delta$&$-\delta$&0\\ \hline
\end{tabular}
\caption{Assignment of family charges under $U(1)_H$.} \label{t21}
\end{center}
\end{table}

\begin{table}
\begin{center}
\begin{tabular}{|c|c|c|c|c|c|c|c|c|c|c|c|c|}\hline
\multicolumn{1}{|c|}{quantum}&\multicolumn{2}{|c|}{Class I}&
\multicolumn{6}{|c|}{Class II}&\multicolumn{2}{|c|}{Class I}&
\multicolumn{2}{|c|}{Class II}\\
number&$\phi_1$&$\phi_2$&$\phi_3$&$\phi_4$&$\phi_5$&$\phi_6$&$\phi_7$&$\phi_8$
&$\phi_9$&$\phi_{10}$&$\phi_{11}$&$\phi_{12}$\\ \hline\hline
H&0&$-\delta$&0&$\Delta$&0&$\delta$&0&$\delta$&$\Delta$&0&$\delta$&0\\
Y&1&0&$-\frac{2}{3}$&$-\frac{2}{3}$&$\frac{4}{3}$&$\frac{4}{3}$&
$-\frac{8}{3}$&$-\frac{8}{3}$&2&2&4&4\\
T&$\frac{1}{2}$&0&1&1&0&0&0&0&1&1&0&0\\
C&1&1&$\overline{6}$&$\overline{6}$&$\overline{6}$&$\overline{6}$&
$\overline{6}$&$\overline{6}$&1&1&1&1\\ \hline
\end{tabular}
\caption{Assignment of charges for scalar fields under $ U(1)_H
\otimes SU(3)_C \otimes SU(2)_L \otimes U(1)_Y $.} \label{t22}
\end{center}
\end{table}

\vspace{3mm} The particle content of the model is such that we can
implement a hierarchical mass generation mechanism, where the
third family of charged fermions obtain mass at tree level, while
the light charged fermions get masses at one and two loops,
respectively. In particular, the radiative mass generation for the
light charged leptons involve the introduction of two $SU(2)_L$
weak scalar triplets with neutral fields, which we allow to get
vacuum expectation values (VEVs). The VEVs of these triplets
contribute very litle to the $W$ and $Z$ masses and simultaneously
allow the generation of tree level Majorana mass terms for the
left-handed neutrinos.

The scalar fields introduced in the model are then divided into
two classes. Class I ( II ) contains scalar fields which acquire
(do not acquire) VEV. These scalar fields are as given in Table 2.

The Yukawa couplings are classified in two types, Dirac(D) and
Majorana(M)[Figs. (1a) and (1b)], respectively,

\begin{equation}
{\cal L}_Y = {\cal L}_{YD} + {\cal L}_{YM}\:,
\end{equation}

\noindent where

\begin{equation}
{\cal
L}_{YD}=Y^u\bar{q}_{L3}\tilde{\phi}_1u_{R3}+Y^d\bar{q}_{L3}\phi_1d_{R3}
+Y^e \bar{l}_{L3}\phi_1\tau_{R3} + H.c\:,
\end{equation}

\vspace{3mm} \noindent with $\bar{\Psi}= \Psi^\dag\gamma^0$,
$\tilde{\phi}=i\sigma_2\phi^*$, $Y^i$, where $i=u,\;d,\;e$ are
coupling constants, and

\vspace{3mm}

\be \begin{array}{lll} {\cal L}_{YM} & =
 & {Y^q}_{12} \;{q^{\alpha}_{1L}}^TC\phi_{3\{\alpha,\beta\}}q^{\beta}_{2L} +
 {Y^q}_{23} \;{q^{\alpha}_{2L}}^TC\phi_{4\{\alpha,\beta\}}q^{\beta}_{3L} +
{Y^q}_{33} \;{q^{\alpha}_{3L}}^TC\phi_{3\{\alpha,\beta\}}q^{\beta}_{3L}  \\
  &  & + \; {Y^u}_{12} \;{u_{1R}}^TC\phi_{7}u_{2R}
+ {Y^u}_{23} \;{u_{2R}}^TC\phi_{8}u_{3R}+ {Y^u}_{33} \;{u_{3R}}^TC\phi_{7}u_{3R}  \\
 &  & + \; {Y^d}_{12} \;{d_{1R}}^TC\phi_{5}d_{2R}
+ {Y^d}_{23} \;{d_{2R}}^TC\phi_{6}d_{3R}+ {Y^d}_{33}\;
{d_{3R}}^TC\phi_{5}d_{3R} + h.c.
\end{array} \ee

\vspace{3mm}

\be \begin{array}{lll} &  & + \; Y_{12}\;
{l^{\alpha}_{1L}}^TC\phi_{10\{\alpha,\beta\}}l^{\beta}_{2L} +
Y_{23}\;{l^{\alpha}_{2L}}^TC\phi_{9\{\alpha,\beta\}}l^{\beta}_{3L}+
Y_{33} \;{l^{\alpha}_{3L}}^TC\phi_{10\{\alpha,\beta\}}l^{\beta}_{3L}  \\
 &  & + \; Y_{12} \;{e_R}^TC\phi_{12}\mu_{R} + Y_{23}\;
{\mu_{R}}^TC\phi_{11}\tau_{R}+ Y_{33}
\;{\tau_{R}}^TC\phi_{12}\tau_{R} +h.c.  \end{array} \ee

\vspace{3mm} \noindent In these couplings $C$ represents the
charge conjugation matrix, $\alpha$ and $\beta$ are weak isospin
indices, and color indices have been omitted. Couplings of Eq.(6)
are introduced for the quark sector, while those of Eq.(7) are
needed for the lepton sector.

\vspace{5mm} \noindent Notice that $\phi_3$ and $\phi_9$ are
represented as

\vspace{3mm}

\be \phi_3 =\left( \begin{array}{cc} {\phi}^{-4/3}& {\phi}^{-1/3} \\
{\phi}^{-1/3}&{\phi}^{2/3} \end{array} \right) , \;\mbox{and} \;\;
\phi_9= \left( \begin{array}{cc} \phi^0& \phi^+ \\ \phi^+ &
\phi^{++} \end{array} \right) \;, \ee

\vspace{3mm}

\noindent where the superscripts denote the electric charge of the
fields ( and corresponding expressions for $\phi_4$ and
$\phi_{10}$ ).

\vspace{3mm}

The most general scalar potential is written as

$$-V(\phi_i)=\sum_{i}{\mu^{2}_{i}{\|\phi_i\|}^2}+
\sum_{i,j}{\lambda_{ij}{\|\phi_i\|}^2{\|\phi_j\|}^2}+\eta_{31}\phi^{\dag}_{1}\phi^{\dag}_{3}
\phi_{3}\phi_{1}+\tilde{\eta}_{31}\tilde{\phi}^{\dag}_{1}\phi^{\dag}_{3}\phi_{3}\tilde{\phi_{1}}$$
$$+\eta_{41}\phi^{\dag}_{1}\phi^{\dag}_{4}\phi_{4}\phi_{1}+\tilde{\eta}_{41}
\tilde{\phi}^{\dag}_{1}\phi^{\dag}_{4}\phi_{4}\tilde{\phi}_{1}+\kappa_{91}
\phi^{\dag}_{1}\phi^{\dag}_{9}\phi_{9}\phi_{1}+\tilde{\kappa}_{91}\tilde{\phi}^{\dag}_{1}
\phi^{\dag}_{9}\phi_{9}\tilde{\phi}_{1}$$
$$+\kappa_{10,1}\phi^{\dag}_{1}\phi^{\dag}_{10}\phi_{10}\phi_{1}+
\tilde{\kappa}_{10,1}\tilde{\phi}^{\dag}_{1}\phi^{\dag}_{10}\phi_{10}
\tilde{\phi}_{1}+\sum_{i\neq{ji},j\neq{1,2}}{\eta_{ij}{\|{\phi_i}^{\dag}\phi_j\|}^2}$$
$$+ \;(\; \rho_1{\phi}^{\dag}_5\phi_6\phi_2+\rho_2{\phi}^{\dag}_7\phi_8\phi_2 +
\lambda_1{\phi}^{\dag}_5{\phi}^\alpha_1\phi_{3\{\alpha,\beta\}}{\phi}^\beta_1+
\lambda_2{\phi}^{\dag}_7{\phi}^\alpha_1\phi_{3\{\alpha,\beta\}}{\phi}^\beta_1$$
$$+\lambda_3Tr(\phi^{\dag}_{3}\phi_4)\phi^{2}_{2}+\lambda_4\phi_5\phi_6
\phi_7\phi_2+\lambda_5\phi_5\phi^{\dag}_{6}\phi^{\dag}_{7}\phi_8+\lambda_6
\phi_2\phi_8\phi^{2}_{5}+y_1\phi_{12}^{\dag}\phi_{11}\phi_2$$
$$+\zeta_1\phi^{2}_{12}{\phi}^\alpha_1\phi_{10\{\alpha,\beta\}}
{\phi}^\beta_1+Y_rTr(\phi^{\dag}_{10}\phi_9)\phi^{2}_{2}+\epsilon_1\phi_5
\phi^{\dag}_{6}\phi^{\dag}_{12}\phi_{11}$$
\begin{equation}
+\epsilon_2\phi^{\dag}_{7}\phi_8\phi_{12}\phi^{\dag}_{11}+h.c. \;)
\end{equation}

\noindent where $Tr$ means trace and in
${\|\phi_i\|}^2=\phi^{+}_{i}\phi_i$ an appropriate contraction of
the $SU(2)_L$ and $SU(3)_C$ indices is understood. The gauge
invariance of this potential requires the relation $\Delta = 2
\delta$ to be hold.

\vspace{3mm}
 In order to break the symmetry, the VEVs for the class I scalar fields
 are assumed to be in the form

\bn \langle\phi_1\rangle = \frac{1}{\sqrt{2}} \left(
\begin{array}{c} 0 \\ v_1  \end{array} \right)
&,& \langle\phi_2\rangle=v_2, \\
\langle\phi_{9}\rangle= \left(
\begin{array}{cc} v_9 & 0 \\ 0&0 \end{array} \right) &,&
\langle\phi_{10}\rangle= \left( \begin{array}{cc} v_{10}&0 \\
0&0 \end{array} \right)\: . \en

\vspace{3mm} \noindent $\langle \phi_1 \rangle$ and $\langle \phi_2
\rangle$ achieve the symmetry breaking sequence

\begin{equation}
U(1)_H \otimes
G_{SM}\stackrel{\langle\phi_{2}\rangle}\longrightarrow
\-G_{SM}\stackrel{\langle\phi_{1}\rangle}\longrightarrow\-SU(3)_C\otimes\-U(1)_Q\:,
\end{equation}

\noindent while the VEVs $v_9$ and $v_{10}$ are extremely small in
order to be consistent with the experimental bounds on the $\rho$
parameter. $M_W=\frac{1}{2} g v_1$ with $v_1\approx 246 \;GeV$,
and I assume $v_2$ in the $TeV$ region.

\vspace{5mm}

The scalar field mixing arises after spontaneous symmetry breaking
(SSB) from the terms in the potential that couple two different
class II fields to one class I field. After SSB the mass matrix
for the scalar fields of charge $+2 \:,(\phi_9, \phi_{10},
\phi_{12}, \phi_{11})$ may be written as

\vspace{3mm}
\begin{equation}
M_{+2}^2= \left( \begin{array}{cccc} e_9^2&Y_r^* v_2^2&0&0 \\ Y_r
v_2^2&e_{10}^2&\zeta_1^* v_1^2\over2&0 \\ 0&\zeta_1
v_1^2\over2&e_{12}^2&y_1 v_2 \\ 0&0&y_1^* v_2&e_{11}^2
\end{array} \right) \;, \label{mmu1}
\end{equation}

\noindent where $e_i^2 = \mu_i^2 + \lambda_{i1}\frac{v_1^2}{2} +
\lambda_{i2} v_2^2$ for $i=9,10,11,12$, and analogous ones for the
$-\frac{4}{3}$ and $\frac{2}{3}$ scalar sectors.

\section{Masses and mixing for charged fermions}

\vspace{3mm}

 Now I give a brief description of the hierarchical mass
generation mechanism for the charged fermions. After the $SSB$ of
the electroweak symmetry down to $U(1)_Q$ of QED, the Yukawa
couplings of Eq.(5) generate tree level masses for the top and
bottom quarks and the $\tau$ lepton. For the light charged
fermions, the scalar fields introduced in the model allow the one
and two loop diagrams of Fig. 2 for the charged lepton mass matrix
elements, and similar ones for the up and down quark sectors. In
the diagrams of Fig. 2 the cross in the internal fermion lines
means tree level mixing and the black dot means one loop mixing.
The diagrams of Figs. 3(a) and 3(b) should be added to the matrix
elements (1,3) and (3,1), respectively.

In the one loop contribution to the mass matrices for the
different charged fermion sectors only the third family of
fermions appears in the internal lines. This generate a rank 2
matrix, which once diagonalized gives the mass eigenstates at this
approximation. Then using these mass eigenstates the next order
contribution is computed, obtaining a matrix of rank 3. After the
diagonalization of this last matrix the mass eigenvalues and
eigenstates are obtained.

\vspace{5mm}

\subsection{Charged leptons}

The nonvanishing contributions from the one loop diagrams of Fig.
2 to the mass terms $\overline{e}_{iR}e_{jL}{\Sigma_{ij}}^{(1)}$
are

\begin{equation}
\Sigma^{(1)}_{22}=m^0_{\tau}\frac{{Y_{23}}^2}{16\pi^{2}}
\sum_{k}{U_{1k}U_{4k}f(M_k,m^0_{\tau})},
\end{equation}

\begin{equation}
\Sigma^{(1)}_{23}=m^0_{\tau}\frac{Y_{23} Y_{33}}{16\pi^{2}}
\sum_{k}{U_{2k}U_{4k}f(M_k,m^0_{\tau})},
\end{equation}

\begin{equation}
\Sigma^{(1)}_{32}=m^0_{\tau}\frac{Y_{23} Y_{33}}{16\pi^{2}}
\sum_{k}{U_{1k}U_{3k}f(M_k,m^0_{\tau})}\:,
\end{equation}

\noindent where $m^0_{\tau}$ is the tree level contribution, $U$ is
the orthogonal matrix which diagonalizes $M_{\phi}^2$, with

\begin{equation}
\Phi_i=U_{ij}\sigma_j \quad , \quad \Phi_1 \equiv \phi_9 \quad ,
\quad \Phi_2 \equiv \phi_{10} \quad , \quad \Phi_3 \equiv \phi_{12}
\quad , \quad \Phi_4 \equiv \phi_{11}
\end{equation}

\noindent being the relation between gauge and mass scalar
eigenfields $\sigma_i$, $i,j=1,2,3,4$, $M_k^2$ are the scalar mass
eigenvalues, and

\begin{equation}
f(a,b) \equiv \frac{a^2}{a^2-b^2}\ln{\frac{a^2}{b^2}} \;.
\end{equation}

\vspace{3mm} \noindent Thus, the one loop contribution to fermion
masses may be written as

\begin{equation}
M^{(1)}_{e}= \left( \begin{array}{ccc} 0&0&0 \\
0&\Sigma^{(1)}_{22}&\Sigma^{(1)}_{23} \\
0&\Sigma^{(1)}_{32}&m^0_{\tau} \end{array} \right)   \equiv \left(
\begin{array}{ccc} 0&0&0 \\ 0& a_2 & a_{23} \\ 0& a_{32}& a_3 \end{array}
\right)\:. \end{equation}

\vspace{5mm} \noindent
 Now, $M_e^{(1)}$ is diagonalized by a biunitary
transformation

\be {V_R^{(1)}}^{\dag} M_e^{(1)} V_L^{(1)} = M_D^{(1)} \;, \ee

\vspace{5mm}

\begin{equation}
{M^{(1)}_{e}}^T M^{(1)}_{e}= \left( \begin{array}{ccc} 0&0&0 \\ 0&
( a_2^2 +a_{32}^2)&( a_2a_{23}+ a_3a_{32}) \\ 0& ( a_2a_{23}+
a_3a_{32})& ( a_3^2 +a_{23}^2) \end{array} \right) \equiv \left(
\begin{array}{ccc} 0&0&0 \\ 0& a_L^{\prime} &c_L^{\prime} \\ 0& c_L^{\prime} &
b_L^{\prime} \end{array} \right)   \;,
\end{equation}

\vspace{5mm}

\begin{equation}
M^{(1)}_{e} {M^{(1)}_{e}}^T= \left( \begin{array}{ccc} 0&0&0 \\ 0&
( a_2^2 +a_{23}^2)&( a_2a_{32}+ a_3a_{23}) \\ 0& ( a_2a_{32}+
a_3a_{23})& ( a_3^2 +a_{32}^2) \end{array} \right) \equiv \left(
\begin{array}{ccc} 0&0&0 \\ 0& a_R^{\prime} &c_R^{\prime} \\ 0& c_R^{\prime} &
b_R^{\prime} \end{array} \right)\:,
\end{equation}

\noindent where in this report (ignoring CP violation), the
orthogonal matrices $V_L^{(1)}$ and $V_R^{(1)}$ are given by
\footnote{I assume the signs: $a_2 > 0 $, $a_{23} < 0$ and $a_{32}
< 0$}

\vspace{3mm}

\be V_L^{(1)} = \left( \begin{array}{ccc} 1 & 0 & 0 \\ 0 &
\cos{\alpha_L} & - \sin{\alpha_L} \\0 & \sin{\alpha_L} &
\cos{\alpha_L}
\end{array} \right) \qquad , \qquad
V_R^{(1)} = \left( \begin{array}{ccc} 1 & 0 & 0 \\ 0 &
\cos{\alpha_R} & - \sin{\alpha_R} \\0 & \sin{\alpha_R} &
\cos{\alpha_R}
\end{array} \right)\:,
\ee

\noindent where

\be \begin{array}{lcl} \cos{\alpha_L} = \alpha^{\prime} (\lambda_+ -
a_L^{\prime}) = \sqrt[]{\frac{\lambda_+ - a_L^{\prime}}{\lambda_+ -
\lambda_-}} & , & \cos{\alpha_R} = \beta^{\prime} (\lambda_+ -
a_R^{\prime}) = \sqrt[]{\frac{\lambda_+ - a_R^{\prime}}{\lambda_+ -
\lambda_-}} \;,                                       \\
                                                      \\
 \sin{\alpha_L} = - \alpha^{\prime} c_L^{\prime} =
\sqrt[]{\frac{\lambda_+ - b_L^{\prime}}{\lambda_+ - \lambda_-}} & ,
& \sin{\alpha_R} = - \beta^{\prime} c_R^{\prime} =
\sqrt[]{\frac{\lambda_+ - b_R^{\prime}}{\lambda_+ - \lambda_-}} \;,
\end{array} \ee

\vspace{3mm} \noindent $\lambda_-$ and $\lambda_+$ are the
solutions of the equation

\be \lambda^2 - B^{\prime} \lambda + D^{\prime} = 0\;, \ee

\bn
 \lambda_{-} = \frac{1}{2} \left[ B^{\prime} -
\sqrt{{B^{\prime}}^2 - 4 D^{\prime}} \; \right] & , & \lambda_{+} =
\frac{1}{2} \left[ B^{\prime} + \sqrt{{B^{\prime}}^2 - 4 D^{\prime}}
\; \right] \;, \en

 \vspace{3mm}

 \be \begin{array}{c} B^{\prime}= a_L^{\prime}+ b_L^{\prime} = a_R^{\prime}+
b_R^{\prime}= a_2^2 + a_3^2 + a_{23}^2 + a_{32}^2= \lambda_- +
\lambda_+  \quad,                                           \\
                                                            \\
 D^{\prime} = a_L^{\prime}b_L^{\prime}- {c_L^{\prime}}^2 =
 a_R^{\prime}b_R^{\prime}- {c_R^{\prime}}^2 = {(a_2a_3 - a_{23}a_{32})}^2=
 \lambda_-\lambda_+ \;,                                     \\
                                                            \\
 a_2a_3 - a_{23}a_{32} > 0 \;,\end{array} \ee

\noindent and

\be \alpha^{\prime} \equiv \frac{1}{\sqrt[]{{c_L^{\prime}}^2 +
{(\lambda_+ - a_L^{\prime})}^2}} \;\;,\;\;
 \beta^{\prime} \equiv \frac{1}{\sqrt[]{{c_R^{\prime}}^2 +
{(\lambda_+ - a_R^{\prime})}^2}} \:,\ee

\vspace{5mm}

 \be
{V_L^{(1)}}^T {M^{(1)}_{e}}^T M^{(1)}_{e} V_L^{(1)} =
{V_R^{(1)}}^T M^{(1)}_{e} {M^{(1)}_{e}}^T V_R^{(1)} = \left(
\begin{array}{ccc} 0&0&0 \\ 0&\lambda_- &0 \\ 0&0&\lambda_+ \end{array}
\right) \:,\ee

\vspace{3mm}

 \be
{V_R^{(1)}}^T M^{(1)}_{e} V_L^{(1)} = \left(
\begin{array}{ccc} 0&0&0 \\ 0&\sqrt[]{\lambda_-} &0\\
0&0&\sqrt[]{\lambda_+} \end{array} \right). \ee

\vspace{5mm} \noindent Therefore, from Eq. (30), up to one loop
level

\be m_1^{(1)}=0 \;\;,\;\; m_2^{(1)}= \sqrt[]{\lambda_-}
\;\;,\;\;m_3^{(1)}= \sqrt[]{\lambda_+} \;,\ee

\vspace{3mm} \noindent with the expected hierarchy $\lambda_- \ll
\lambda_+ $.

\vspace{5mm}

\noindent Two loop contributions for charged leptons:

\begin{equation}
\Sigma_{11}^{(2)} =\frac{Y_{12}^2}{16\pi^{2}} \sum_{k,i}
m_i^{(1)}(V_L^{(1)})_{2i}(V_R^{(1)})_{2i} U_{2k}U_{3k} f(M_k,
m_i^{(1)}),
\end{equation}

\vspace{2mm}

\begin{equation}
\Sigma_{12}^{(2)} =\frac{Y_{12} Y_{23}}{16\pi^{2}} \sum_{k,i}
m_i^{(1)}(V_L^{(1)})_{3i}(V_R^{(1)})_{2i} U_{1k}U_{3k} f(M_k,
m_i^{(1)}),
\end{equation}

\vspace{2mm}

\be \begin{array}{lll}\Sigma_{13}^{(2)} & = & \frac{Y_{12}
Y_{23}}{16\pi^{2}} \sum_{k,i}
m_i^{(1)}(V_L^{(1)})_{2i}(V_R^{(1)})_{2i} U_{1k}U_{3k} f(M_k,
m_i^{(1)})                                   \\
                                             \\
& & + \frac{Y_{12} Y_{33}}{16\pi^{2}} \sum_{k,i}
m_i^{(1)}(V_L^{(1)})_{3i}(V_R^{(1)})_{2i} U_{2k} U_{3k} f(M_k,
m_i^{(1)})                             \;, \end{array} \ee

\vspace{2mm}

\begin{equation}
\Sigma_{21}^{(2)} =\frac{Y_{12} Y_{23}}{16\pi^{2}} \sum_{k,i}
m_i^{(1)}(V_L^{(1)})_{2i}(V_R^{(1)})_{3i} U_{2k}U_{4k} f(M_k,
m_i^{(1)}),
\end{equation}

\vspace{2mm}

\be \begin{array}{lll} \Sigma_{31}^{(2)} & = & \frac{Y_{12}
Y_{23}}{16\pi^{2}} \sum_{k,i}
m_i^{(1)}(V_L^{(1)})_{2i}(V_R^{(1)})_{2i} U_{2k}U_{4k} f(M_k,
m_i^{(1)})                                  \\
                                            \\
  & & + \frac{Y_{12} Y_{33}}{16\pi^{2}} \sum_{k,i}
m_i^{(1)}(V_L^{(1)})_{2i}(V_R^{(1)})_{3i} U_{2k} U_{3k} f(M_k,
m_i^{(1)})                               \;, \end{array} \ee

\noindent where $i=2,3$ and $k=1,2,3,4$.

\vspace{3mm}

Hence the mass matrix for charged leptons up to two loop
contributions may be written in good approximation as

\vspace{3mm}

 \be M^{(2)}_{e} \approx \left(
\begin{array}{ccc}
\Sigma^{(2)}_{11}&\Sigma^{(2)}_{12}&\Sigma^{(2)}_{13}\\
\Sigma^{(2)}_{21}& \sqrt[]{\lambda_-}&0\\
\Sigma^{(2)}_{31}&0&\sqrt[]{\lambda_+} \end{array} \right) \equiv
\left( \begin{array}{ccc} \Sigma_{11}&\Sigma_{12}&\Sigma_{13}\\
\Sigma_{21}& \sqrt[]{\lambda_-}&0\\
\Sigma_{31}&0&\sqrt[]{\lambda_+} \end{array} \right). \ee

\vspace{3mm}

\noindent In the limit $M_k >> m_{\tau}^0$ the function $f(a,b)$
behaves as $\ln{\frac{a^2}{b^2}}$. In this limit, and introducing
the $m_{i}^{(1)}$ one loop mass eigenvalues, Eq.(31), the orthogonal
matrices $V_L^{(1)}$ and $V_R^{(1)}$, Eq.(23), the relationships

\vspace{3mm}

\large \be \begin{array}{ccc} \cos\alpha_L \cos\alpha_R  = \frac{a_3
\sqrt{\lambda_+} - a_2 \sqrt{\lambda_-}}{\lambda_+ - \lambda_-} & ,
& \cos\alpha_L \sin\alpha_R  = - \: \frac{a_{23} \sqrt{\lambda_+} +
a_{32} \sqrt{\lambda_-}}{\lambda_+ - \lambda_-}        \;,      \\
                                                             \\
\sin\alpha_L \sin\alpha_R  =  \frac{a_2 \sqrt{\lambda_+} - a_3
\sqrt{\lambda_-}}{\lambda_+ - \lambda_-}   & , & \sin\alpha_L
\cos\alpha_R  = - \: \frac{a_{32} \sqrt{\lambda_+} + a_{23}
\sqrt{\lambda_-}}{\lambda_+ - \lambda_-}  \;,
\end{array} \ee

\normalsize

\vspace{3mm} \noindent and using the orthogonality of $U$, one
obtains explicitly

\vspace{5mm}

\large
\be \begin{array}{lll} \Sigma_{11}=a_2 \: \sigma > 0 &,&  \\
                                                   \\
\Sigma_{12} = \Sigma_{21}= \frac{1}{c_{\alpha}} \: \frac{a_{23} \: a_{32}}{a_3} > 0 &,&
                      c_{\alpha} \equiv \frac{Y_{33}}{Y_{12}} \\
                                                        \\
 \Sigma_{13}= c_{\alpha} \: a_{23} \: \sigma + \frac{1}{c_{\alpha}} \: \frac{a_{2} \: a_{32}}{a_3} < 0 &,& \\
                                                          \\
 \Sigma_{31}= c_{\alpha} \: a_{32} \: \sigma + \frac{1}{c_{\alpha}} \frac{a_{2} \: a_{23}}{a_3} < 0 &,&
 \end{array} \normalsize \ee

\normalsize \vspace{3mm} \noindent where the parameter $\sigma$ is
defined as

\vspace{3mm}

\begin{equation}
\sigma \equiv \frac{{Y_{12}}^2}{16\pi^{2}} \sum_{k}{U_{2k}U_{3k}
\ln{\frac{M_k^2}{m_\tau^0}}} > 0\;.\end{equation}

\subsection{Quarks}

\vspace{5mm}

Performing an analogous analysis, the one loop contributions for
the down quark sector are

\vspace{3mm}
\begin{equation}
\left( \Sigma^{(1)}_{22} \right)^d=m^0_{b}\frac{Y_{23}^q
Y_{23}^d}{16\pi^{2}} \sum_{k}{U_{1k}^d U_{4k}^d f(M_k^d,m_b^0)}
\equiv a_2^d \:,
\end{equation}

\vspace{3mm}

\begin{equation}
\left( \Sigma^{(1)}_{23}\right)^d=m^0_{b}\frac{Y_{33}^q
Y_{23}^d}{16\pi^{2}} \sum_{k}{U_{2k}^d U_{4k}^d f(M_k^d,m_b^0)}
\equiv a_{23}^d \:,
\end{equation}

\vspace{3mm}

\begin{equation}
\left( \Sigma^{(1)}_{32}\right)^d=m^0_{b}\frac{Y_{23}^q
Y_{33}^d}{16\pi^{2}} \sum_{k}{U_{1k}^d U_{3k}^d f(M_k^d,m_b^0)}
\equiv a_{32}^d\:,
\end{equation}

\vspace{3mm} \noindent where $m^0_{b}$ is the tree level
contribution, $U^d$ is the orthogonal matrix which diagonalizes
$M_{d \:\phi}^2$, with

\begin{equation}
\Phi_i^d=U_{ij}^d \sigma_j^d \quad , \quad \Phi_1^d \equiv \phi_4
\quad , \quad \Phi_2^d \equiv \phi_{3} \quad , \quad \Phi_3^d \equiv
\phi_{5} \quad , \quad \Phi_4^d \equiv \phi_{6}
\end{equation}

\noindent being the relation between gauge and mass scalar
eigenfields $\sigma_i^d$, $i,j=1,2,3,4$ and ${M_k^d}^2$ are the mass
eigenvalues.

\vspace{3mm} \noindent Similarly, the two loop contributions for
the down quark sector may be expressed as

\vspace{3mm}

\Large
\be \begin{array}{lll} \Sigma_{11}^d & = & a_2^d \: \sigma^d > 0 \:,  \\
                                                   \\
\Sigma_{12}^d & = & \frac{Y_{12}^d}{Y_{33}^d} \: \frac{a_{23}^d \:
a_{32}^d}{m_b^0} > 0  \:,                                \\
                                                        \\
\Sigma_{21}^d & = & \frac{Y_{12}^q}{Y_{33}^q} \: \frac{a_{23}^d \:
a_{32}^d}{m_b^0} > 0  \:,                                \\
                                                          \\
 \Sigma_{13}^d & = & \frac{Y_{33}^q}{Y_{12}^q} \: a_{23}^d \: \sigma^d + \frac{Y_{12}^d}{Y_{33}^d} \:
 \frac{a_{2}^d \: a_{32}^d}{m_b^0} < 0 \:, \\
                                                          \\
 \Sigma_{31}^d & = & \frac{Y_{33}^d}{Y_{12}^d} \: a_{32}^d \: \sigma^d + \frac{Y_{12}^q}{Y_{33}^q} \:
 \frac{a_{2}^d \: a_{23}^d}{m_b^0} < 0 \:,

 \end{array} \normalsize \ee

\normalsize \vspace{3mm} \noindent with $\sigma^d$  defined as

\begin{equation}
\sigma^d = \frac{Y_{12}^q Y_{12}^d}{16\pi^{2}} \sum_{k}{U_{2k}^d
U_{3k}^d \ln{\frac{{M_k^d}^2}{m_b^0}}} > 0\;,\end{equation}

\vspace{3mm} \noindent and analogous ones for the up quark sector.

\vspace{5mm} Notice that the radiative corrections give rise to the
following hierarchy among the parameters in the mass matrices for
charged leptons

\vspace{3mm}
\be \Sigma_{11}\quad ,\quad \Sigma_{12} \quad , \quad
\Sigma_{21} \quad , \quad |\Sigma_{13}| \quad , \quad |\Sigma_{31}|
\quad \ll \quad a_2 \quad , \quad |a_{23}| \quad , \quad |a_{32}|
\quad \ll \quad a_3=m_\tau^o \;, \ee

\vspace{3mm} \noindent and similarly for the u and d quark sectors.

\vspace{3mm}
 The matrix $M_e^{(2)}$, Eq. (37), is
diagonalized by a new biunitary transformation

\be {V_R^{(2)}}^{\dag} M_e^{(2)} V_L^{(2)} = M_D^{(2)} \;. \ee

\vspace{3mm}

In this case, with the aid of results of the Appendix, the
orthogonal matrix $V_L^{(2)}$ may be written as

\Large

\be V_L^{(2)} = \left( \begin{array}{ccc} \sqrt[]{\frac{(b_L -
\lambda_1) (c_L - \lambda_1) - f_L^2}{(\lambda_2 -
\lambda_1)(\lambda_3 - \lambda_1)}}& \sqrt[]{\frac{(b_L -
\lambda_2) (c_L - \lambda_2) - f_L^2}{(\lambda_1 -
\lambda_2)(\lambda_3 - \lambda_2)}} & - \;\: \sqrt[]{\frac{(b_L -
\lambda_3) (c_L - \lambda_3) - f_L^2}{(\lambda_1 -
\lambda_3)(\lambda_2 - \lambda_3)}} \\
        &                       &                       \\ - \;\;
\sqrt[]{\frac{(a_L - \lambda_1) (c_L - \lambda_1) -
e_L^2}{(\lambda_2 - \lambda_1)(\lambda_3 - \lambda_1)}} &
\sqrt[]{\frac{(a_L - \lambda_2) (c_L - \lambda_2) -
e_L^2}{(\lambda_1 - \lambda_2)(\lambda_3 - \lambda_2)}} & - \;\;
\sqrt[]{\frac{(a_L - \lambda_3) (c_L - \lambda_3) -
e_L^2}{(\lambda_1 - \lambda_3)(\lambda_2 - \lambda_3)}} \\
        &                       &                       \\
\sqrt[]{\frac{(a_L - \lambda_1) (b_L - \lambda_1) -
d_L^2}{(\lambda_2 - \lambda_1)(\lambda_3 - \lambda_1)}} &
\sqrt[]{\frac{(a_L - \lambda_2) (b_L - \lambda_2) -
d_L^2}{(\lambda_1 - \lambda_2)(\lambda_3 - \lambda_2)}} &
\sqrt[]{\frac{(a_L - \lambda_3) (b_L - \lambda_3) -
d_L^2}{(\lambda_1 - \lambda_3)(\lambda_2 - \lambda_3)}}
\end{array} \right) \;, \normalsize \ee

\normalsize

\vspace{3mm}

\noindent where $\lambda_i \;, i = 1,2,3 $ are the eigenvalues of
${M^{(2)}_{e}}^T M^{(2)}_{e}$ ( $M^{(2)}_{e} {M^{(2)}_{e}}^T$);
$\lambda_1 \equiv m_e^2$, $\lambda_2 \equiv m_\mu^2$, and
$\lambda_3 \equiv m_\tau^2$ for charged leptons.
 A similar expression is obtained for $V_R^{(2)}$ by replacing
L by R, where the L and R parameters are defined as

\begin{equation}
{M^{(2)}_{e}}^T M^{(2)}_{e} \equiv \left(
\begin{array}{ccc} a_L&d_L&e_L \\ d_L& b_L &f_L \\ e_L& f_L & c_L
\end{array} \right) \;\;,\;\; M^{(2)}_{e} {M^{(2)}_{e}}^T \equiv
\left( \begin{array}{ccc} a_R&d_R&e_R\\ d_R& b_R &f_R \\ e_R& f_R
& c_R \end{array} \right) \:,
\end{equation}

\vspace{3mm} \noindent

\bn \begin{array}{lll} a_L=\Sigma_{11}^2 + \Sigma_{21}^2 +
\Sigma_{31}^2 \;,& d_L=\Sigma_{21}\sqrt{\lambda_-} +
\Sigma_{11}\Sigma_{12} \;,&
e_L=\Sigma_{31}\sqrt{\lambda_+} + \Sigma_{11}\Sigma_{13} \\
                                                                \\
& b_L=\lambda_- + \Sigma_{12}^2 \;,& f_L=\Sigma_{12}\Sigma_{13} \\
                                                                \\
&                 & c_L=\lambda_+ + \Sigma_{13}^2 \end{array} \;,\en

\vspace{3mm}

\bn \begin{array}{lll} a_R=\Sigma_{11}^2 + \Sigma_{12}^2 +
\Sigma_{13}^2 \;,& d_R=\Sigma_{12}\sqrt{\lambda_-} +
\Sigma_{11}\Sigma_{21} \;,&
e_R=\Sigma_{13}\sqrt{\lambda_+} + \Sigma_{11}\Sigma_{31} \\
                                                                \\
& b_R=\lambda_- + \Sigma_{21}^2 \;,& f_R=\Sigma_{21}\Sigma_{31} \\
                                                                \\
&                 & c_R=\lambda_+ + \Sigma_{31}^2 \end{array} \;,
\en

\vspace{3mm} \noindent and similarly for the u and d quark sectors.

\vspace{3mm} Up to now I have realized a complete exact analytical
diagonalization of the resulting charged fermion mass matrices at
one and two loops contributions. Thus, from Eqs. (23) and (49) one
may obtain exact expressions for the orthogonal matrices

\be V_L \equiv V_L^{(1)} V_L^{(2)} \qquad , \qquad V_R \equiv
V_R^{(1)} V_R^{(2)} \:.\ee

\vspace{3mm} \noindent In particular, the $V_{CKM}$ quark mixing
matrix (ignoring CP violation) takes the form

\be V_{CKM} = ( V_{uL}^{(1)} V_{uL}^{(2)})^{T} V_{dL}^{(1)}
V_{dL}^{(2)} \ee

\subsection{Approximate mixing matrices for charged fermions}

\vspace{3mm} Taking advantage of the strong hierarchy of masses
observed in the quarks and charged leptons, it is possible to make
good approximations for the mixing matrices $V_L^{(1)}$ and
$V_L^{(2)}$ given in the Eqs. (23) and (49), respectively. For
instance, the tree level contribution defines the magnitude of the
masses for the heaviest fermions $m_t\:, m_b$, and $m_\tau$ in
each sector. One loop contribution determines the masses for
$m_c\:, m_s$, and $m_\mu$ and the mixing angle between the 2 and 3
families, giving simultaneously small corrections to the masses of
the heaviest fermions. Finally, the two loop contribution gives
masses to the lightest fermions $u, d$ and $e$ and determines
their mixing with the families 2 and 3, giving some tiny
corrections to the masses of the heavier fermions. I use this
perturbative mechanism to make the following approximations:

\bn \lambda_- \ll \lambda_+ & \mbox{implies} &
\frac{D^\prime}{{B^\prime}^2}=\frac{\lambda_-
\lambda_+}{(\lambda_- + \lambda_+)^2}= \frac{\lambda_-}{\lambda_+}
\frac{1}{(1 + \frac{\lambda_-}{\lambda_+})^2} \ll 1 \:,\en

\noindent and expanding now the square root in Eq.(26), one gets

\large \be \begin{array}{lcl} \lambda_- = \frac{B^\prime}{2}
\left[1- \sqrt{1- 4 \frac{D^\prime}{{B^\prime}^2}} \right] &
\approx & \frac{D^\prime}{B^\prime} \approx \frac{(a_2a_3 -
a_{23}a_{32})^2}{a_3^2}                            \\
& = & \left(a_2 - \frac{a_{23}a_{32}}{a_3} \right)^2= (a_2 -
\Sigma_{12})^2 \;, \end{array} \normalsize \ee

\normalsize
\vspace{3mm}

\noindent and hence from Eqs. (21) and (27), $\lambda_+ -
b^\prime_L=a^\prime_L - \lambda_- \approx a_2^2 + a_{32}^2 - (a_2
- \Sigma_{12})^2 \thickapprox a_{32}^2 $. Thus, from Eqs.
(23)-(27), the strong hierarchy of masses leads to the
approximation

\be V_L^{(1)} \approx \left(
\begin{array}{ccc} 1 & 0 & 0 \\
0 & c_{23} & - s_{23} \\ 0 & s_{23} & c_{23}
\end{array} \right) \quad , \quad
s_{23} \equiv \sqrt{\frac{a_{32}^2}{\lambda_3 - \lambda_2}} \quad
, \quad c_{23}= \sqrt{1- s_{23}^2} \:,\ee

\vspace{3mm}

\noindent  where I have set $\lambda_+ - \lambda_- \approx
\lambda_3 - \lambda_2$. Then, in this approach the main
contribution to the mixing angle between the families 2 and 3
comes from the mixing matrix element $(V_L^{(1)})_{23}$. So,
according to this perturbative approach the two loop contribution
$(V_L^{(2)})_{23}$ is negligible, that is, $|(V_L^{(2)})_{23}| \ll
|(V_L^{(1)})_{23}|$, and hence it is a good approach to set
$(V_L^{(2)})_{23} \approx 0$  in Eq.(49). Let me discuss further
about the consistence of this approximation.

\vspace{3mm}

From the results of the Appendix, by replacing $a, b, c, d, e, f
\rightarrow a_L, b_L, c_L, d_L, e_L, f_L$, one gets the exact
equation

\be \Delta_{2L}(\lambda_3)\equiv \lambda_3^2 -(a_L + c_L)\lambda_3 +
a_L c_L - e_L^2=(a_L c_L - e_L^2 - \lambda_1 \lambda_3) -
\lambda_3(\lambda_2 - b_L) \:, \ee

\vspace{3mm} \noindent where $a_L c_L - e_L^2=(\Sigma_{11}^2 +
\Sigma_{21}^2)\lambda_{+} - 2
\Sigma_{11}\Sigma_{13}\Sigma_{31}\sqrt{\lambda_{+}} +
(\Sigma_{21}^2 + \Sigma_{31}^2)\Sigma_{13}^2 \sim O(2
\;\mbox{loops})^2 \; \lambda_+$. Let me recall here that from
Eq.(31), $\lambda_2 \rightarrow \lambda_{-}$ and $\lambda_3
\rightarrow \lambda_{+}$ if one ignores two loop contributions to
the masses for the 2 and 3 families. Therefore, because $\lambda_1
\sim O(2 \;\mbox{loops})^2$ and $\lambda_2 - b_L = \lambda_2 -
\lambda_{-} - \Sigma_{12}^2 \sim O(2\;\mbox{loops})^2$ by
construction, one concludes that the right-hand side of Eq. (58)
is $\sim O(2\;\mbox{loops})^2 \; \lambda_3$, and hence

\vspace{3mm}

\be
|(V_L^{(2)})_{23}|=\sqrt{\frac{\Delta_{2L}(\lambda_3)}{h(\lambda_3)}}
\sim O(\frac{2\;\mbox{loops}}{\sqrt{\lambda_3}}) \lesssim
O(\frac{m_e}{m_\tau}) \ll 1\:. \ee

\vspace{3mm}

\noindent One can also directly write

\be  \begin{array}{lll} \lambda_3^2 -(a_L + c_L)\lambda_3 + a_L c_L
- e_L^2 & = & \lambda_3^2 - [\lambda_{-} + \lambda_{+} + a_L +
\Sigma_{13}^2]\lambda_3 + [\lambda_{-} + \frac{a_Lc_L -
e_L^2}{\lambda_3}]\lambda_3                              \\
                                                         \\
& \approx & \lambda_3^2 - [\lambda_{-} + \lambda_{+}]\lambda_3 +
\lambda_{-} \lambda_3 \end{array} \:.\ee

\vspace{3mm}

\noindent Comparing now Eq. (60) with  Eq. (25), one concludes, in
what concern the magnitudes for the heaviest fermion in each
charged sector; $m_t^2$, $m_b^2$, and $m_{\tau}^2$, that the
eigenvalue $\lambda_3$ satisfies with good approximation the
quadratic equation

\be \Delta_{2L}(\lambda_3)\equiv \lambda_3^2 -(a_L + c_L)\lambda_3
+ a_L c_L - e_L^2 \approx 0 \:,\ee

\vspace{3mm} \noindent and hence $(V_L^{(2)})_{23} \approx 0$ is
in good agreement with the radiative corrections. Using now Eq.
(61) and the orthogonality of $V_L^{(2)}$, Eq.(49), one reaches
the approximation

\vspace{3mm}

\be V_L^{(2)} \approx \left( \begin{array}{ccc} c_{12} c_{13} &
c_{13} s_{12} & - s_{13} \\ - s_{12} & c_{12} & 0 \\ c_{12} s_{13} &
s_{12} s_{13} & c_{13} \end{array} \right) = \left(
\begin{array}{ccc} c_{13} & 0 & - s_{13} \\
0 & 1 & 0 \\ s_{13} & 0 & c_{13}
\end{array} \right)\:\left(
\begin{array}{ccc} c_{12} & s_{12} & 0 \\
- s_{12} & c_{12} & 0 \\ 0 & 0 & 1
\end{array} \right)
\:, \ee

\vspace{3mm}

\noindent where the $s_{12}$ and $s_{13}$ mixing angles are
identified as

\vspace{3mm}

\large \be  \begin{array}{lcl} s_{12} \equiv
\sqrt{\frac{\Sigma_{21}^2}{\lambda_2 - \lambda_1}} &,&
s_{13} \equiv \sqrt{\frac{\Sigma_{31}^2}{\lambda_3 - \lambda_1}} \\
                                                   \\
c_{12}= \sqrt{1-s_{12}^2} &,& c_{13}= \sqrt{1-s_{13}^2}\:,
\end{array} \normalsize \ee

\normalsize

\vspace{3mm} \noindent and therefore

\vspace{3mm}

\large \be V_L \equiv V^{(1)}_L V^{(2)}_L \approx \left(
\begin{array}{ccc}
 c_{12} c_{13} & c_{13} s_{12} & - s_{13} \\
 -c_{23} s_{12} - c_{12} s_{13} s_{23} & c_{12} c_{23} - s_{12} s_{13} s_{23} & - c_{13} s_{23} \\
c_{12} c_{23} s_{13} - s_{12} s_{23} & c_{12}s_{23} +
c_{23}s_{12}s_{13} & c_{13} c_{23}
\end{array} \right)_L\:. \normalsize \ee

\normalsize

\vspace{3mm}

It is important to emphasize here that Eq. (64) it is not a
parametrization but an approximation that is consistent with

\begin{itemize}
\item the strong hierarchy of masses for quarks and charged
leptons and \item the radiative corrections.
\end{itemize}

\vspace{3mm}

A similar analysis for the R handed mixing matrices yields

\vspace{3mm}

\large \be V_R \equiv V^{(1)}_R V^{(2)}_R \approx \left(
\begin{array}{ccc}
 c^{\prime}_{12} c^{\prime}_{13} & c^{\prime}_{13} s^{\prime}_{12} & - s^{\prime}_{13} \\
 -c^{\prime}_{23} s^{\prime}_{12} - c^{\prime}_{12} s^{\prime}_{13} s^{\prime}_{23} &
 c^{\prime}_{12} c^{\prime}_{23} - s^{\prime}_{12} s^{\prime}_{13} s^{\prime}_{23} & - c^{\prime}_{13}s^{\prime}_{23} \\
-
s^{\prime}_{12}s^{\prime}_{23}+c^{\prime}_{12}c^{\prime}_{23}s^{\prime}_{13}
& c^{\prime}_{12}s^{\prime}_{23} +
c^{\prime}_{23}s^{\prime}_{12}s^{\prime}_{13} & c^{\prime}_{13}
c^{\prime}_{23}
\end{array} \right)_R\:, \normalsize \ee

\normalsize \noindent where

\vspace{3mm}

\large \be s^{\prime}_{12} \equiv
\sqrt{\frac{\Sigma_{12}^2}{\lambda_2 - \lambda_1}}\quad , \quad
s^{\prime}_{13} \equiv \sqrt{\frac{\Sigma_{13}^2}{\lambda_3 -
\lambda_1}} \quad , \quad s^{\prime}_{23} \equiv
\sqrt{\frac{a_{23}^2}{\lambda_3 - \lambda_2}} \normalsize \;.\ee

\normalsize
\vspace{5mm}

So, the strong hierarchy of masses for charged leptons leads to the
approach

\vspace{3mm}

\large \be \begin{array}{lcl} s_{12}  \approx
\frac{\Sigma_{21}}{m_\mu} \thickapprox  O\left(\frac{2 \;
\mbox{loops}}{1 \; \mbox{loop}}\right) & , & s_{12}^{\prime}
\approx \frac{\Sigma_{12}}{m_\mu}  \thickapprox  O\left(\frac{2 \;
\mbox{loops}}{1 \; \mbox{loop}}\right)              \;,   \\
                                               \\
s_{23} \approx  \frac{|a_{32}|}{m_\tau}  \thickapprox O\left(\frac{1
\; \mbox{loop}}{\mbox{tree level}}\right) & , & s_{23}^{\prime}
\approx \frac{|a_{23}|}{m_\tau}  \thickapprox
O\left(\frac{1 \; \mbox{loop}}{\mbox{tree level}}\right) \;,     \\
                                               \\
s_{13}  \approx  \frac{|\Sigma_{31}|}{m_\tau}  \thickapprox
O\left(\frac{2 \; \mbox{loops}}{\mbox{tree level}}\right) & , &
s_{13}^{\prime}  \approx  \frac{|\Sigma_{13}|}{m_\tau} \thickapprox
O\left(\frac{2 \; \mbox{loops}}{\mbox{tree level}}\right) \;.
\end{array} \normalsize \ee

\normalsize \vspace{3mm} Hence one concludes that the radiative mass
mechanism naturally yields the hierarchy

\large \be  \begin{array}{ccc} s_{13} \; < \; s_{12} \quad,\quad
s_{23} \qquad & ; & \qquad s_{13}^{\prime} \; < \; s_{12}^{\prime}
\quad , \quad s_{23}^{\prime} \:.
\end{array} \normalsize \ee

\normalsize

\vspace{3mm} Notice that the approximations realized in this
section, Eqs. (57) and (62)-(68), may equally be applied to the
e,u, and d charged sectors by replacing properly the mass
parameters: $m_{\tau}^0 \rightarrow m_t^0 \:,\:m_b^0$ \quad ,
\quad $a_2,a_{23},a_{32} \rightarrow
a_2^{u,d},a_{23}^{u,d},a_{32}^{u,d}$ and
$\Sigma_{11},\Sigma_{12},\Sigma_{13},\Sigma_{21},\Sigma_{31}
\rightarrow
\Sigma_{11}^{u,d},\Sigma_{12}^{u,d},\Sigma_{13}^{u,d},\Sigma_{21}^{u,d},\Sigma_{31}^{u,d}$,
respectively, in an obvious notation.

\pagebreak

\section{Quantitative analysis of masses and mixing for charged leptons}

Using the orthogonality conditions of U and Eq.(171) of the
Appendix, one can compute the parameters involved in the charged
lepton mass matrices; one gets explicitly

\vspace{3mm}

\large \be \begin{array}{lcl} a_2 & = & m_{\tau}^0
\frac{{Y_{23}}^2}{16 {\pi}^2} \sum_{k=1}^4 U_{1k} U_{4k}
\ln{\frac{M_k^2}{{m_\tau^0}^2}} = m_{\tau}^0 \frac{{Y_{23}}^2}{16
{\pi}^2}
{\sum_{k=1}^4\frac{f_4(\eta_k)}{h(\eta_k)}\ln{\frac{\eta_k}{{m_\tau^0}^2}}}
\equiv m_{\tau}^0 \frac{{Y_{23}}^2}{16 {\pi}^2}\,F_{22} \\
                                                       \\
 a_{23}  & = & m_{\tau}^0 \frac{Y_{23} Y_{33}}{16 {\pi}^2}
\sum_{k=1}^4 U_{2k} U_{4k} \ln{\frac{M_k^2}{{m_\tau^0}^2}} =
m_{\tau}^0 \frac{Y_{23} Y_{33}}{16 {\pi}^2} \sum_{k=1}^4
\frac{g_4(\eta_k)}{h(\eta_k)} \ln{\frac{\eta_k}{{m_\tau^0}^2}}
\equiv - m_{\tau}^0 \frac{ Y_{23} Y_{33}}{16 {\pi}^2}\,F_{23}\\
                                                            \\
a_{32}  & = & m_{\tau}^0 \frac{Y_{23} Y_{33}}{16 {\pi}^2}
\sum_{k=1}^4 U_{1k} U_{3k} \ln{\frac{M_k^2}{{m_\tau^0}^2}} =
m_{\tau}^0 \frac{Y_{23} Y_{33}}{16 {\pi}^2} \sum_{k=1}^4
\frac{f_3(\eta_k)}{h('eta_k)} \ln{\frac{\eta_k}{{m_\tau^0}^2}}
\equiv - m_{\tau}^0
\frac{Y_{23} Y_{33}}{16 {\pi}^2}\,F_{32}      \\
                                              \\
\sigma & = & \frac{Y_{12}^2}{16 {\pi}^2} \sum_{k=1}^4 U_{2k} U_{3k}
\ln{\frac{M_k^2}{{m_\tau^0}^2}} = \frac{Y_{12}^2}{16 {\pi}^2}
\sum_{k=1}^4 \frac{g_3(\eta_k)}{h(\eta_k)}
\ln{\frac{\eta_k}{{m_\tau^0}^2}} \equiv \frac{{Y_{12}}^2}{16
{\pi}^2}\,F_{\sigma}
\end{array} \;, \normalsize \ee

\normalsize

\vspace{3mm} \noindent where $\eta_k \equiv M_k^2$, and the
dimensionless functions $F_{22}$, $F_{23}$, $F_{32}$, and
$F_{\sigma}$ may be written as

\vspace{3mm}
\large \be \begin{array}{lllll} F_{22} & = &
\sum_{k=2}^4 \frac{f_4(\eta_k)}{h(\eta_k)}
ln{\frac{\eta_k}{\eta_1}} & = & \frac{bcd}{|h(\eta_2)|}
\ln{\frac{\eta_2}{\eta_1}} - \frac{bcd}{h(\eta_3)}
\ln{\frac{\eta_3}{\eta_1}} +
\frac{bcd}{|h(\eta_4)|} \ln{\frac{\eta_4}{\eta_1}}  >  0   \\
                                            \\
F_{23} & = & - \sum_{k=2}^4 \frac{g_4(\eta_k)}{h(\eta_k)}
ln{\frac{\eta_k}{\eta_1}} & = &  \frac{cd(a_1 -
\eta_2)}{|h(\eta_2)|} \ln{\frac{\eta_2}{\eta_1}} - \frac{cd(a_1 -
\eta_3)}{h(\eta_3)} \ln{\frac{\eta_3}{\eta_1}} +
\frac{cd(a_1 - \eta_4}{|h(\eta_4)|} \ln{\frac{\eta_4}{\eta_1}}  >  0    \\
                                            \\
F_{32} & = & - \sum_{k=2}^4 \frac{f_3(\eta_k)}{h(\eta_k)}
ln{\frac{\eta_k}{\eta_1}} & = &
   \frac{bc(a_4 - \eta_2)}{|h(\eta_2)|} \ln{\frac{\eta_2}{\eta_1}}
- \frac{bc(a_4 - \eta_3)}{h(\eta_3)} \ln{\frac{\eta_3}{\eta_1}} +
\frac{bc(a_4 - \eta_4)}{|h(\eta_4)|} \ln{\frac{\eta_4}{\eta_1}} >  0    \\
                                       \\
F_{\sigma} & = & \sum_{k=2}^4 \frac{g_3(\eta_k)}{h(\eta_k)}
ln{\frac{\eta_k}{\eta_1}} & = &   \frac{c(a_1 - \eta_2)(a_4 -
\eta_2)}{|h(\eta_2)|} \ln{\frac{\eta_2}{\eta_1}} - \frac{c(a_1 -
\eta_3)(a_4 - \eta_3)}{h(\eta_3)} \ln{\frac{\eta_3}{\eta_1}} \\
                                                             \\
   &  &         &   & +
\frac{c(a_1 - \eta_4)(a_4 - \eta_4)}{|h(\eta_4)|}
\ln{\frac{\eta_4}{\eta_1}}   >  0
 \end{array}  \;, \normalsize \ee

\normalsize \vspace{3mm} \noindent where

\be |h(\eta_2)| \equiv (\eta_2 - \eta_1)(\eta_3 - \eta_2)(\eta_4 -
\eta_2) \qquad , \qquad |h(\eta_4)| \equiv (\eta_4 - \eta_1)(\eta_4
- \eta_2)(\eta_4 - \eta_3) \,.\ee

\vspace{5mm}

\noindent To leading order in the radiative loop corrections, one
reaches the approximations

\vspace{3mm}

\large \be \begin{array}{l} m_\tau \equiv \sqrt{\lambda_3} \approx
\sqrt{\lambda_+} \approx a_3 = m_{\tau}^0              \;,      \\
                                                              \\
m_\mu \equiv \sqrt{\lambda_2} \approx \sqrt{\lambda_-} \approx a_2 \;,    \\
                                                              \\
m_e \equiv \sqrt{\lambda_1} \approx \Sigma_{11} = a_2 \:\sigma
\approx m_\mu \:\sigma                             \;,
\end{array} \normalsize \ee

\normalsize

\vspace{3mm} \noindent and hence from Eqs. (69)-(72):

\Large \be \begin{array}{lcr} \frac{a_2}{m_\tau^0} =
\frac{{Y_{23}}^2}{16 {\pi}^2}\:F_{22} \approx \frac{m_\mu}{m_\tau}
& ; & \sigma = \frac{Y_{12}^2}{16 {\pi}^2}\;F_{\sigma} \approx
\frac{m_e}{m_\mu}                                      \\
\hspace{-1cm} \mbox{or}                                              \\
\frac{Y_{23}}{4 \pi} \approx \frac{.243842}{\sqrt{F_{22}}} & ; &
\frac{Y_{12}}{4 \pi} \approx \frac{.0695437}{\sqrt{F_{\sigma}}}\:.
\end{array} \normalsize \ee

\normalsize \vspace{3mm}

\noindent So, in this approach the following relationships hold:

\be \left( \frac{Y_{12} Y_{23}}{16 \pi^2} \right)^2 \approx
\frac{\frac{m_e}{m_\tau}}{F_{22} \:F_{\sigma}} = \frac{2.875643
\:\mbox{x}\: 10^{-4}}{F_{22} \:F_{\sigma}} \;, \ee

\vspace{3mm}

\be \frac{Y_{12}}{Y_{23}} \thickapprox (.285199)
\frac{\sqrt{F_{22}}}{\sqrt{F_{\sigma}}} \;, \ee

\vspace{3mm}

\Large
\be \begin{array}{lll} \frac{|a_{23}|}{m_{\mu}}
\thickapprox \frac{|a_{23}|}{a_2} = \frac{Y_{33} F_{23}}{Y_{23}
F_{22}} = c_{\alpha} \frac{Y_{12} F_{23}}{Y_{23} F_{22}} =
c_{\alpha}
(.285199) \frac{F_{23}}{\sqrt{F_{22} F_{\sigma}}}         \\
                                                             \\
\frac{|a_{32}|}{m_{\mu}} \thickapprox \frac{|a_{32}|}{a_2} =
\frac{Y_{33} F_{32}}{Y_{23} F_{22}} = c_{\alpha} \frac{Y_{12}
F_{32}}{Y_{32} F_{22}} = c_{\alpha} (.285199)
\frac{F_{32}}{\sqrt{F_{22} F_{\sigma}}}
\end{array} \normalsize \;, \ee

\normalsize

\vspace{5mm}

\Large \be \Sigma_{12}=\Sigma_{21}= \frac{1}{c_{\alpha}}
\frac{a_{23} a_{32}}{a_3}= \frac{1}{c_{\alpha}} \frac{|a_{23}|
|a_{32}|}{m_{\tau}^0} \thickapprox  \frac{1}{c_{\alpha}} |a_{23}|
\:s_{23}^e = \frac{1}{c_{\alpha}} |a_{32}| \:s_{23}^{\prime e}
\normalsize  \;, \ee

\normalsize \vspace{5mm}

\Large \be \begin{array}{c} |\Sigma_{31}| = c_{\alpha} |a_{32}|
\sigma + \frac{1}{c_{\alpha}} \frac{a_2 |a_{23}|}{a_3} \thickapprox
c_{\alpha} \frac{m_e}{m_\mu} |a_{32}| +  \frac{1}{c_{\alpha}}
\frac{m_\mu}{m_\tau} |a_{23}|                    \\
                                                 \\
 |\Sigma_{13}| = c_{\alpha} |a_{23}| \sigma + \frac{1}{c_{\alpha}}
\frac{a_2 |a_{32}|}{a_3} \thickapprox c_{\alpha} \frac{m_e}{m_\mu}
|a_{23}| +  \frac{1}{c_{\alpha}} \frac{m_\mu}{m_\tau} |a_{32}|
\end{array} \normalsize \;,\ee

\normalsize \vspace{3mm} \noindent where the superscript $e$
denotes the charged lepton sector. Therefore, the mixing angles in
$(V_{eL})$, Eq. (64), and $(V_{eR})$, Eq. (65), may be expressed
as

\vspace{3mm}

\Large \be \begin{array}{lll} s_{23}^e& \approx &
\sqrt{\frac{a_{32}^2}{\lambda_3 - \lambda_2}} \approx
\frac{|a_{32}|}{m_\tau} = c_{\alpha} \left( \frac{Y_{12}
Y_{23}}{16 \pi^2} \right)\:F_{32} = c_{\alpha}(0.016957) \frac{F_{32}}{\sqrt{F_{22} \,F_{\sigma}}}\\
                                                  \\
 s_{12}^e & \approx & \sqrt{\frac{\Sigma_{21}^2}{\lambda_2 -
 \lambda_1}} \approx \frac{\Sigma_{21}}{m_\mu} = \frac{1}{c_{\alpha}}
 \frac{|a_{23}|}{m_\mu} \:s_{23}^e \approx (.285199) \frac{F_{23}}{\sqrt{F_{22} \,F_{\sigma}}} \:s_{23}^e     \\
                                                   \\
 s_{13}^e & \approx & \sqrt{\frac{\Sigma_{31}^2}{\lambda_3 -
 \lambda_1}} \approx \frac{|\Sigma_{31}|}{m_\tau} \approx
  c_{\alpha} \frac{m_e}{m_\mu} \:s_{23}^e +
 (\frac{m_\mu}{m_\tau})^2 \:\frac{s_{12}^e}{s_{23}^e}
\end{array} \;, \normalsize \ee

\normalsize
\vspace{4mm} \noindent and

\Large \be \begin{array}{lll} s_{23}^{\prime e}& \approx &
\sqrt{\frac{a_{23}^2}{\lambda_3 - \lambda_2}} \approx
\frac{|a_{23}|}{m_\tau} = c_{\alpha} \left( \frac{Y_{12}
Y_{23}}{16 \pi^2} \right)\:F_{23} = c_{\alpha} (.016957) \frac{F_{23}}{\sqrt{F_{22} \,F_{\sigma}}} \\
                                                  \\
 s_{12}^{\prime e} & \approx & \sqrt{\frac{\Sigma_{12}^2}{\lambda_2 -
 \lambda_1}} \approx \frac{\Sigma_{12}}{m_\mu} = \frac{1}{c_{\alpha}}
 \frac{|a_{32}|}{m_\mu} \:s_{23}^{\prime e} \approx (.285199) \frac{F_{32}}{\sqrt{F_{22} \,F_{\sigma}}}
 \:s_{23}^{\prime e}                             \\
                                                   \\
 s_{13}^{\prime e} & \approx & \sqrt{\frac{\Sigma_{13}^2}{\lambda_3 -
 \lambda_1}} \approx \frac{|\Sigma_{13}|}{m_\tau} \approx
 c_{\alpha} \frac{m_e}{m_\mu} \:s_{23}^{\prime e} + (\frac{m_\mu}{m_\tau})^2
\: \frac{s_{12}^{\prime e}}{s_{23}^{\prime}}
\end{array} \;, \normalsize \ee

\normalsize

\vspace{3mm} \noindent respectively, with the relations

\be  s_{12}^e = s_{12}^{\prime e} \qquad \text{and hence} \qquad
\frac{s_{23}^e}{s_{23}^{\prime}} = \frac{|a_{32}|}{|a_{23}|} =
\frac{F_{32}}{F_{23}} \:. \ee

\subsection{Upper bounds for charged lepton mixing angles}

\vspace{3mm}

Each particular set of scalar mass parameters in Eq.(165) define a
spectrum of scalar mass eigenvalues $\eta_1$, $\eta_2$, $\eta_3$,
$\eta_4$, the values for $F_{22}$, $F_{23}$, $F_{32}$ and
$F_{\sigma}$ through the Eq.(70), as well as the magnitudes for
mixing angles in $V_{eL}$ and $V_{eR}$ through Eqs. (79) and (80).
A numerical evaluation shows that the variation of these mixing
angles is relatively small for a large region in the space mass
parameters. So, in order to find out the orders of magnitude for
these mixing angles, let me redefine the parameters of
$M_{\phi}^2$, Eq.(165), in such a way

\be M_{\phi}^2 =\left( \begin{array}{cccc} a_1^{\prime} &
b^{\prime}
& 0 & 0\\ b^{\prime} & a_2^{\prime} & c^{\prime} & 0\\ 0 & c^{\prime} & a_3^{\prime} & d^{\prime}\\
0 &  0 & d^{\prime} & a_4^{\prime}
\end{array} \right) M^2
\;,\ee

\vspace{3mm} \noindent where $a_1^{\prime}$, $a_2^{\prime}$,
$a_3^{\prime}$, $a_4^{\prime}$, $b^{\prime}$, $c^{\prime}$ and
$d^{\prime}$ are positive real numbers, while $M$ is a mass
parameter in the TeV region. Mixing angles in $V_{eL}$ and
$V_{eR}$ do not depend on $M$. The value of $M^2$ may be
determined for instance by specifying the value of the lightest
scalar mass eigenvalue $\eta_1 \equiv M_1^2$.

\vspace{3mm}

Setting for example and simplicity $a_1^{\prime}=a_4^{\prime}$ ,
$b^{\prime}=c^{\prime}=d^{\prime}$ one gets $F_{23}=F_{32}$ and
hence $s_{23}^e = s_{23}^{\prime e}$, $s_{12}^e = s_{12}^{\prime
e}$, $s_{13}^e = s_{13}^{\prime e}$. So, in the simplified parameter
space region defined by $a_1^{\prime} = a_4^{\prime}=1$, \quad $2
\leqq a_2^{\prime} \leqq 120$, \quad $3 \leqq a_3^{\prime} \leqq
125$, \quad $1 \leqq b^{\prime}= c^{\prime}= d^{\prime} \leqq 10$,
one gets the following range of values for mixing angles in the
charged lepton sector:

\vspace{3mm}

\large \be \begin{array}{rcl} 4.666163 \times {10}^{-3} \lesssim &
s_{23}^e=s_{23}^{\prime e} & \lesssim 1.34417 \times {10}^{-2}    \;,      \\
                                                      \\
0.105688 \lesssim & \frac{s_{12}^e}{s_{23}^e} = \frac{s_{12}^{\prime e}}{s_{23}^{\prime e}} & \lesssim 0.240805 \;,  \\
                                                      \\
4.931603 \times {10}^{-4} \lesssim & s_{12}^e= s_{12}^{\prime e}
& \lesssim 3.236832 \times {10}^{-3}               \;,    \\
                                                      \\
3.904076 \times {10}^{-4} \lesssim & s_{13}^e=s_{13}^{\prime e} &
\lesssim 9.123709 \times {10}^{-4} \;,
\end{array} \normalsize \ee

\normalsize

\vspace{3mm} \noindent where the upper and lower bounds are
obtained with the values  $a_2^{\prime}=2$, $a_3^{\prime}=3$,
$b^{\prime}= c^{\prime}= d^{\prime}=1$, and $a_2^{\prime}=120$,
$a_3^{\prime}=125$, $b^{\prime}= c^{\prime}= d^{\prime}=10$,
respectively, and where I have used the range of values
$c_{\alpha}= \frac{Y_{33}}{Y_{12}}=\frac{c}{2 d}=0.742528 -
0.938792$ corresponding to the global parameter space region
defined by Eq. (103) in the analysis of neutrino mixing.

\pagebreak

\section{Neutrino masses and $V_{PMNS}$}

\vspace{5mm}

From the Yukawa couplings of Eq.(7), the mass matrix for the
left-handed neutrinos is obtained as

\begin{equation}
M_{\nu}= \left( \begin{array}{ccc} 0& Y_{12} \:v_{10}/2 &0\\
Y_{12} \:v_{10}/2 & 0 & Y_{23} \:v_9/2\\ 0& Y_{23} \:v_9/2& Y_{33}
\:v_{10}
\end{array} \right) \equiv \left(
\begin{array}{ccc}0& d &0\\ d & 0 & f\\ 0& f& c \end{array} \right)
\:. \end{equation}

\noindent One may diagonalize $M_\nu$ as

\be U_\nu^T M_\nu U_\nu = M_\nu^d \;, \ee

\noindent where $M_\nu^d \equiv diag (\xi_1 , \xi_2 , \xi_3 )$ is
the diagonal matrix with $\xi_1$ , $\xi_2$, and $\xi_3$ being the
eigenvalues of $M_\nu$, and $U_\nu$ is the rotation matrix which
connects the gauge states with the corresponding eigenstates.

\vspace{3mm}

The eigenvalues $\xi_1$, $\xi_2$, and $\xi_3$ satisfy the
following nonlinear relationships with the parameters $d$, $f$ and
$c$ of $M_\nu$, Eq.(84):

\be \begin{array}{rcl}\xi_1 + \xi_2 + \xi_3 & = & c \\
    \xi_1\xi_2 + \xi_1\xi_3 + \xi_2\xi_3 & = & - d^2 - f^2 \\
    \xi_1\xi_2\xi_3 & = & - d^2 c
\end{array} \ee

The square matrix elements  ${{U_\nu}}_{ij}^2$ may be obtained
from those of  $({V_L^{(2)}})_{ij}^2$, Eq.(49), by replacing\\
$a_L, b_L, c_L, d_L, e_L, f_L \rightarrow 0, 0, c, d, 0, f$ and
$\lambda_i \rightarrow \xi_i$ respectively. However, from the
Eq.(86) and assuming $c, d, f>0$, it is easy to conclude that one
of the $\xi_i, \;i=1,2,3$ is negative. Thus, the eigenvalues
$\xi_i$ cannot be directly associated to the physical neutrino
masses.

\vspace{3mm}

 Setting $\xi_3>0$ and computing explicitly the ${{U_\nu}}_{ij}^2$
 elements, one arrives to the following statements:

\vspace{3mm}

 1. Assuming normal hierarchy
   \be \xi_1^2 < \xi_2^2 < \xi_3^2 \;\; \mbox{implies} \;\; \xi_1>0 \;\;
    \mbox{and} \;\;\xi_2<0 \;,
   \ee

2. Assuming the hierarchy
 \be \xi_2^2 < \xi_1^2 < \xi_3^2 \;\; \mbox{implies}\;\; \xi_1<0 \;\;
 \mbox{and}\;\; \xi_2>0
   \ee

In what follows, I assume a normal hierarchy for the squared
eigenvalues $\xi_i^2$ as in Eq.(87)\footnote{The second
possibility, Eq.(88), does not change any conclusion about this
model}. In the literature there exists a lot of models dealing
with normal neutrino mass hierarchy\cite{numodels}. Now I define
the unitary matrix $V_\nu \equiv U_\nu \;
diag(1,i,1)$\cite{bookmoha}, which may be written as

\vspace{3mm}

\Large
\be V_\nu = \left( \begin{array}{ccc} \sqrt{\frac{f^2 \:m_2
m_3}{(m_2^2 - m_1^2)(m_3^2 - m_1^2)}}& - i\; \sqrt{\frac{f^2 \:m_1
m_3}{(m_2^2 - m_1^2)(m_3^2 - m_2^2)}} &  \;\: \sqrt{\frac{f^2
\:m_1 m_2}{(m_3^2 - m_1^2)(m_3^2 - m_2^2)}} \\
 &              &                           \\
 \;\; \sqrt{\frac{f^2 \:c \:m_1}{(m_2^2 - m_1^2)(m_3^2 - m_1^2)}}
 & i\; \sqrt{\frac{f^2 \:c \:m_2}{(m_2^2 - m_1^2)(m_3^2 - m_2^2)}} &
\;\; \sqrt{\frac{f^2 \:c \:m_3}{(m_3^2 - m_1^2)(m_3^2 - m_2^2)}} \\
  &             &                                            \\
-\;\sqrt{\frac{(d^2 - m_1^2)(d^2 + f^2 - m_1^2)}{(m_2^2 -
m_1^2)(m_3^2 - m_1^2)}} & - i\; \sqrt{\frac{(m_2^2 - d^2)(d^2 +
f^2 - m_2^2)}{(m_2^2 - m_1^2)(m_3^2 - m_2^2)}} &
\sqrt{\frac{(m_3^2- d^2)(m_3^2 - d^2 - f^2)}{(m_3^2 - m_1^2)(m_3^2
- m_2^2)}}
\end{array} \right) \normalsize \;, \ee

\vspace{3mm}

\normalsize

\noindent or equivalently in the form

\Large \be V_\nu = \left( \begin{array}{ccc} \sqrt[]{\frac{m_2 m_3
(m_3 - m_2)}{c(m_1 + m_2)(m_3 - m_1)}}& - i\; \sqrt[]{\frac{m_1
m_3(m_1+m_3)}{c(m_1 + m_2)(m_3 +
m_2)}} & \;\: \sqrt[]{\frac{m_1 m_2(m_2 - m_1)}{c(m_3 - m_1)(m_3 + m_2)}} \\
 &              &             \\ \;\; \sqrt[]{\frac{m_1(m_3 -
 m_2)}{(m_1 + m_2)(m_3 - m_1)}} & i\; \sqrt[]{\frac{m_2(m_1 +
 m_3)}{(m_1 + m_2)(m_3 + m_2)}} &
 \;\; \sqrt[]{\frac{m_3(m_2 - m_1)}{(m_3 - m_1)(m_3 + m_2)}} \\
  &             &                                            \\
 -\;\sqrt[]{\frac{m_1(m_1 + m_3)(m_2 - m_1)}{c(m_1 + m_2)(m_3 - m_1)}} & - i\;
\sqrt[]{\frac{m_2(m_3 - m_2)(m_2 - m_1)}{c(m_1 + m_2)(m_3 + m_2)}}
& \sqrt[]{\frac{m_3(m_1 + m_3)(m_3 - m_2)}{c(m_3 - m_1)(m_3 +
m_2)}}
\end{array} \right) \normalsize \ee

\vspace{3mm}

\normalsize

\noindent after using properly the results of the Appendix and
making the identification

\be  (\xi_1 , - \:\xi_2 , \xi_3 ) =  (m_1 , m_2 , m_3 )\; \ee

\noindent between the eigenvalues $\xi_i$ and the physical
neutrino masses $m_1$, $m_2$, and $m_3$. Therefore, for neutrinos
the transformation between gauge, ${\psi_\nu^0}_L^T = (\nu_e^0,
\nu_\mu^0, \nu_\tau^0)_L$, and mass eigenstates, ${\psi_\nu}_L^T =
(\nu_1, \nu_2, \nu_3)_L$, is

\vspace{3mm} \be {\psi_\nu^0}_L = V_{\nu}\: {\psi_\nu}_L   \:. \ee

\vspace{4mm} \noindent From Eq.(91) and the definition of $V_\nu$ it
is easy to verify that

\be V_\nu^T M_\nu V_\nu = diag (m_1 , m_2 , m_3 ) \:,\ee

\vspace{3mm} \noindent and hence one may write Eq.(86) in terms of
neutrino masses:

\bn \begin{array}{rcl} m_1 + m_3 - m_2 & = & c     \,, \\
    m_1 m_2 - m_1 m_3 + m_2 m_3 & = &  d^2 + f^2   \;, \\
    m_1 m_2 m_3 & = & d^2 c       \;.
\end{array} \en

\vspace{3mm} \noindent
 The combination of these relationships yields the useful equality

\be f^2 c = (m_3 - m_2)(m_3 + m_1)(m_2 - m_1) \:.\ee

\vspace{3mm}

Notice that Eqs. (94) and (95) allow one to write all the matrix
elements $(V_{\nu})_{ij}$ completely in terms of the physical
neutrino masses $m_1 \:,m_2 \:$, and $m_3$ as in Eq.(90).

\subsection{$V_{PMNS}$ lepton mixing matrix }

The current experimental study of neutrino oscillation phenomena
gives as a result that in the lepton sector the mixing matrix
$V_{PMNS}$ behaves close to the so-called "tribimaximal mixing"
(TBM)\cite{tbm}. In particular, according to Eq.(1), the mixing
angles $\theta_{12}$ and $\theta_{23}$ are large,
$\sin{\theta_{12}}$ and $\sin{\theta_{23}} \lesssim O(1)$, while
$\theta_{13}$ has not yet been measured. So, taking the ranges of
values in Eq.(83) as the typical orders of magnitude for the
mixing angles in the charged lepton sector, it is then clear that
mixing in the lepton sector should come almost completely from
neutrino mixing, and then one may approach with good precision

\vspace{3mm}

\be V_{PMNS}\equiv (V_{eL})^\dag V_\nu \approx V_\nu \:.\ee

\vspace{3mm} \noindent Thus, from Eqs. (89), (90), and (96), the
$V_{PMNS}$ lepton mixing matrix in this model may be approached as

\vspace{3mm}

\large \be V_{PMNS} \approx \left(
\begin{array}{ccc}
 c_{12} c_{13} & - i \: c_{13} s_{12} &  s_{13} \\
 c_{23} s_{12} - c_{12} s_{23} s_{13} & i \:(c_{12} c_{23} + s_{12} s_{23} s_{13}) &  c_{13} s_{23} \\
- \:(s_{12} s_{23} + c_{12} c_{23} s_{13}) & - i \:( c_{12} s_{23}
- c_{23} s_{12} s_{13}) & c_{13} c_{23}
\end{array} \right)\:, \normalsize \ee

\normalsize \vspace{3mm} \noindent where the lepton mixing angles
are identified as

\Large \be \begin{array}{lcl} S_{13}^2 \equiv (V_\nu)_{13}^2 =
\frac{f^2 \:m_1 m_2}{(m_3^2 - m_1^2)(m_3^2 - m_2^2)} & = &
\frac{m_1 m_2(m_2-m_1)}{(m_1 + m_3 - m_2 )(m_3-m_1)(m_3+m_2)}     \;, \\
                                                           \\
 S_{12}^2 \equiv \frac{m_3^2 - m_1^2}{m_2^2 - m_1^2} \frac{f^2
\:m_1 m_3}{(m_3^2 - m_1^2)(m_3^2 - m_2^2) - f^2\:m_1 m_2} & = &
\frac{s_{13}^2}{c_{13}^2} \frac{m_3}{m_2} \frac{m_3^2 - m_1^2}{m_2^2
- m_1^2}                                                   \;, \\
                                                            \\
 S_{23}^2 \equiv  \frac{f^2 \:c m_3}{(m_3^2 - m_1^2)(m_3^2 - m_2^2)
- f^2\:m_1 m_2} & = & \frac{s_{13}^2}{c_{13}^2} \frac{m_3}{m_2}
\frac{c}{m_1}                             \;,
\end{array} \normalsize  \ee

\normalsize
\vspace{3mm} \noindent and

\Large \be \begin{array}{lcl} c_{13}^2 = 1 - s_{13}^2 & = &
\frac{(m_3^2 - m_1^2)(m_3^2 - m_2^2)- f^2 \:m_1 m_2}{(m_3^2 -
m_1^2)(m_3^2 - m_2^2)}                       \;,        \\
                                                     \\
 c_{23}^2 = 1 - s_{23}^2 & = & \frac{1}{c_{13}^2} \frac{m_3}{c}
\frac{(m_3 + m_1)(m_3 - m_2)}{(m_3 - m_1)(m_3 + m_2)} \;,  \\
                                                      \\
 c_{12}^2 = 1 - s_{12}^2 & = & \frac{s_{13}^2}{c_{13}^2} \frac{m_3
(m_3^2 - m_2^2)}{m_1 (m_2^2 - m_1^2)}         \;.
\end{array} \normalsize \ee

\normalsize

\vspace{5mm} \noindent The combination of the last two equations
yields

\Large \be \begin{array}{lll} \sin^2{2\theta_{12}}=4 \frac{m_1
m_2}{(m_1
+ m_2)^2} \frac{(m_3^2 -m_1^2)(m_3^2 - m_2^2)}{(m_3^2 - m_2^2 - m_1^2 + m_1 m_2)^2}           \;,  \\
                                                    \\
\sin^2{2\theta_{23}}=4 \frac{(m_1 + m_3 - m_2)(m_2 - m_1)(m_3 -
m_2)(m_3 + m_1)}{(m_3^2 - m_2^2 - m_1^2 + m_1 m_2)^2} \;.
\end{array} \normalsize \ee

\normalsize
\vspace{3mm}

\subsubsection{Numerical analysis}

It is clear from Eqs. (89), (90), and (97)-(100) that this model
predicts $s_{13}^2
> 0$, implying some deviation from the TBM limit. The allowed range
of values for lepton mixing depends on the value, or range of
values, used for $s_{13}$. To perform a numerical analysis let me
introduce the parameters $k$ and $x$ defined as

\bn k \equiv \frac{m_3 - m_2}{m_1} & , &  x \equiv \frac{m_2}{m_1}
> 1 \;. \en

\vspace{3mm} \noindent One may write all the matrix elements of
$V_{PMNS}$ in terms of these two parameters. In particular
$s_{13}^2$ may be expressed as

\vspace{3mm}
\be s_{13}^2=\frac{x(x-1)}{(k+1)(k+x-1)(k+2x)} \:. \ee

\vspace{3mm} \noindent The last equation may be used now to invert
$x$ in terms of $s_{13}^2$ and k, and thus $\sin^2{2 \theta_{12}}$
and $\sin^2{2 \theta_{23}}$ may be written in terms of $s_{13}^2$
and k. A numerical analysis shows that for the range of values
$0.033 \leqslant s_{13}^2 \leqslant 0.04$ one may obtain large
mixing angles for $\theta_{12}$ and $\theta_{23}$ within the
allowed limits of Eq.(1). This region for $s_{13}^2$ is consistent
with the upper bound provided by the CHOOZ experiment
\cite{s13choozb} ($s_{13}^2 \lesssim 0.04$). I point out the
allowed magnitudes for lepton mixing in the following
$(s_{13}^2,k)$ parameter space regions:

\vspace{3mm}

\be \text{\bf global parameter space:} \qquad \qquad
 0.033 \leqslant s_{13}^2 \leqslant 0.04 \quad , \quad 3.2 \leqslant k \leqslant 4.1
\ee

\vspace{3mm}

\noindent This region yields the range of mixing angles:

\vspace{3mm}
\be                                                 \\
 0.64865 \leqslant \sin^2{2 \theta_{12}} \leqslant 0.818112 \qquad
, \qquad 0.813749 \leqslant \sin^2{2 \theta_{23}} \leqslant 0.919788
\:, \ee

\vspace{2mm} \noindent and the $V_{PMNS}$ unitary mixing matrix with
the following range of magnitudes:

\vspace{2mm}

\be V_{PMNS} \approx \left( \begin{array}{rrl} 0.830485 - 0.874368 &
0.442132 - 0.526588 & 0.181659 - 0.2\\0.254605 - 0.371263 & 0.765592
- 0.773352 & 0.524249 - 0.586563 \\
0.401143 - 0.432288  & 0.367937 - 0.461950 & 0.784821 - 0.831963
\end{array} \right) \:.\ee

\vspace{3mm}

Recall that the above range of values is restricted by the
constraints imposed by the unitarity of $V_{PMNS}$; that is,
choosing a specific value of one entry further restricts the range
of values for the other entries. It is clear from Eq.(104) that
only part of the values in Eq.(105) are within the allowed limits
of Eq.(1). Given a particular value for $s_{13}^2$ in Eq.(103), it
is possible to specify the k parameter region where lepton mixing
lies within these allowed limits. I point out below these range of
values for $s_{13}^2=0.034 \:,\:0.037 \:, \text{and}\: 0.04$,
respectively:

\vspace{5mm}

\be \begin{array}{lrcl} \text{\bf Case A:}
 & s_{13}^2=0.034 & , &  3.88182 \leqslant
k \leqslant 4.02591 \qquad (4.50978 \leqslant x \leqslant
4.79497)                                      \\
                                              \\
& 0.7 \leqslant \sin^2{2 \theta_{12}} \leqslant 0.719315 & , &
0.87 \leqslant \sin^2{2 \theta_{23}} \leqslant 0.878086
\end{array} \:, \ee

\vspace{3mm} \be V_{PMNS} \approx \left( \begin{array}{rrl} 0.859588
- 0.864610 & 0.467386 - 0.476558 & 0.184390 \\
0.298043 - 0.308732 & 0.771902 - 0.772539 &
0.555744 - 0.560673\\
0.404500 - 0.407176 & 0.420784 - 0.429807 & 0.807246 - 0.810647
\end{array} \right) \ee

\vspace{5mm}

\be \begin{array}{lrcl} \text{\bf Case B:} & s_{13}^2=0.037 & , &
3.52059 \leqslant k \leqslant 3.8732 \qquad (4.18727 \leqslant x
\leqslant
4.91525)                                      \\
                                              \\
 & 0.7 \leqslant \sin^2{2 \theta_{12}} \leqslant 0.749742 & ,
& 0.87 \leqslant \sin^2{2 \theta_{23}} \leqslant 0.890806
\end{array} \:, \ee

\vspace{3mm} \be V_{PMNS} \approx \left( \begin{array}{rrl} 0.849926
- 0.863266 & 0.466660 - 0.490536 & 0.192353 \\
0.289950 - 0.318087 & 0.768718 - 0.770414 &
0.554881 - 0.567795\\
0.413159 - 0.420055 & 0.410422 - 0.434385 & 0.800381 - 0.809387
\end{array} \right) \ee

\vspace{5mm}

\be \begin{array}{lrcl} \text{\bf Case C:} & s_{13}^2=0.04 & , &
3.21323 \leqslant k \leqslant 3.73671 \qquad (3.91261 \leqslant x
\leqslant
5.03901)                                      \\
                                              \\
 & 0.7 \leqslant \sin^2{2 \theta_{12}} \leqslant 0.777209 & ,
& 0.87 \leqslant \sin^2{2 \theta_{23}} \leqslant 0.902305
\end{array} \:, \ee

\vspace{3mm} \be V_{PMNS} \approx \left( \begin{array}{rrl} 0.840573
- 0.861920 & 0.465933 - 0.503425 & 0.2 \\
0.282093 - 0.326762 & 0.765697 - 0.768410 &
0.554016 - 0.57443\\
0.421328 - 0.432045 & 0.400339 - 0.438694 & 0.793744 - 0.808125
\end{array} \right) \ee

\vspace{5mm}

An additional analysis shows that for $0.0375 \lesssim s_{13}^2
\lesssim 0.04$ and $0.035 \lesssim s_{13}^2 \lesssim 0.04$ one may
specify a k parameter region where lepton mixing lies within the
$3 \sigma$ allowed ranges reported in
Refs.\cite{numixbggm,numixbgg}, respectively.

\subsubsection{Neutrino masses}

With the purpose to obtain some rough estimation for the order of
magnitudes of neutrino masses let me use the range of values for
lepton mixing in Eqs. (106)-(111) and the bounds for ${\Delta
m}_{\text{sol}}^2$ and ${\Delta m}_{\text{atm}}^2$ of Eq. (1). One
gets the following neutrino masses.

\vspace{4mm}

{\bf ${\Delta m}_{\text{sol}}^2 = m_2^2-m_1^2=(x^2-1)\:m_1^2$:}

\vspace{3mm}
\be \begin{array}{ccc} m_1  & \thickapprox &
\left(\:1.796 - 2.145 \quad , \quad 1.750 - 2.320 \quad , \quad 1.706 - 2.494 \:\right) \times 10^{-3}\:eV \\
                                                                \\
m_2 & \thickapprox &
\left(\:8.103 - 10.28 \quad , \quad 7.331 - 11.40 \quad , \quad 6.675 - 12.56 \:\right) \times 10^{-3}\:eV \\
                                                                \\
m_3 & \thickapprox & \left(\:1.507 - 1.892 \quad , \quad 1.349 -
2.039 \quad , \quad 1.215 - 2.188 \:\right) \times 10^{-2}\:eV
\end{array} \:,\ee

\vspace{3mm} \noindent where the first, second, and third range of
values for each $m_i\:,i=1,2,3$ correspond to $s_{13}^2=0.034$,
$s_{13}^2=0.037$, and $s_{13}^2=0.04$ respectively.

\vspace{5mm} {\bf ${\Delta m}_{\text{atm}}^2 =
m_3^2-m_2^2=k(k+2x)\:m_1^2$:}

\vspace{3mm}
\be \begin{array}{ccc} m_1 & \thickapprox &
\left(\:5.053 - 8.117 \quad , \quad 5.135 - 8.876 \quad , \quad 5.207 - 9.645 \:\right) \times 10^{-3}\:eV \\
                                                                  \\
m_2 & \thickapprox &
\left(\:2.279 - 3.892 \quad , \quad 2.150 - 4.363 \quad , \quad  2.037 - 4.860 \:\right) \times 10^{-2}\:eV \\
                                                                  \\
m_3 & \thickapprox & \left(\:4.240 - 7.160 \quad , \quad 3.958 -
7.801 \quad , \quad  3.710 - 8.464 \:\right) \times 10^{-2}\:eV
\end{array} \ee

\vspace{5mm}

\section{Quantitative analysis of quark masses and $V_{CKM}$}

To leading order in the radiative loop corrections, one gets the
approximations

\vspace{3mm}

\large \be \begin{array}{lcl} m_b \equiv \sqrt{\lambda_3^d} \approx
\sqrt{\lambda_+^d} \approx a_3^d = m_b^0              & , & m_t
\equiv \sqrt{\lambda_3^u} \approx
\sqrt{\lambda_+^u} \approx a_3^u = m_t^0       \\
                                                              \\
m_s \equiv \sqrt{\lambda_2^d} \approx \sqrt{\lambda_-^d} \approx a_2^d & , &
m_c \equiv \sqrt{\lambda_2^u} \approx \sqrt{\lambda_-^u} \approx a_2^u    \\
                                                              \\
m_d \equiv \sqrt{\lambda_1^d} \approx \Sigma_{11}^d = a_2^d
\:\sigma^d \approx m_s \:\sigma^d                             & , &
m_u \equiv \sqrt{\lambda_1^u} \approx \Sigma_{11}^u = a_2^u
\:\sigma^u \approx m_c \:\sigma^u
\end{array} \normalsize \ee

\normalsize

\vspace{3mm} \noindent and then one obtains the relations

\Large \be \begin{array}{lcl} \sigma^d  =  \frac{Y_{12}^q
\:Y_{12}^d}{16 \pi^2}\:F_{\sigma}^d  \approx  \frac{m_d}{m_s} & , &
\sigma^u  =  \frac{Y_{12}^q
\:Y_{12}^u}{16 \pi^2}\:F_{\sigma}^u  \approx  \frac{m_u}{m_c}  \\
                                                           \\
\frac{a_2^d}{m_b^0} =  \frac{Y_{23}^q \:Y_{23}^d}{16
\pi^2}\:F_{22}^d  \approx  \frac{m_s}{m_b} & , & \frac{a_2^u}{m_t^0}
=  \frac{Y_{23}^q \:Y_{23}^u}{16 \pi^2}\:F_{22}^u  \approx
\frac{m_c}{m_t}
\end{array}
\normalsize \ee

\normalsize

\vspace{5mm} \noindent where the functions $F_{22}^{d,u} \:,
F_{23}^{d,u} \:,F_{32}^{d,u} \:,F_{\sigma}^{d,u}$ are defined
analogous to those for the charged lepton sector in Eq.(70).

\vspace{4mm} Hence the mixing angles for the d and u quark
sectors, $V_L^d$ and $V_L^u$, Eq.(64), may be approximated as

\Large \be \begin{array}{lcl} s_{23}^d  \approx
\frac{Y_{33}^d}{Y_{23}^d} \:\frac{F_{32}^d}{F_{22}^d}
\:\frac{m_s}{m_b} & , & s_{23}^u  \approx
\frac{Y_{33}^u}{Y_{23}^u} \:\frac{F_{32}^u}{F_{22}^u} \:\frac{m_c}{m_t} \\
                                                         \\
s_{12}^d \approx
\frac{Y_{12}^q}{Y_{23}^q}\:\frac{F_{23}^d}{F_{22}^d} \:s_{23}^d & ,
&
s_{12}^u  \approx  \frac{Y_{12}^q}{Y_{23}^q}\:\frac{F_{23}^u}{F_{22}^u} \:s_{23}^u \\
                                                         \\
s_{13}^d  \approx  \frac{Y_{33}^d}{Y_{12}^d} \:\frac{m_d}{m_s}
\:s_{23}^d + (\frac{m_s}{m_b})^2 \: \frac{s_{12}^d}{s_{23}^d} & , &
s_{13}^u  \approx  \frac{Y_{33}^u}{Y_{12}^u} \:\frac{m_u}{m_c}
\:s_{23}^u + (\frac{m_c}{m_t})^2 \: \frac{s_{12}^u}{s_{23}^u}
\end{array} \normalsize \ee

\normalsize

\subsection{\bf Numerical analysis}

To explore the allowed magnitudes for mixing angles in $V_{CKM}$
and without lost of generality, let me assume for simplicity the
relationships

\vspace{4mm}

\large \be F_{22}^d=F_{22}^u \equiv {\cal{F}}_{22} \quad , \quad
F_{23}^d=F_{23}^u \equiv {\cal{F}}_{23}\quad , \quad
F_{32}^d=F_{32}^u \equiv {\cal{F}}_{32} \quad , \quad
F_{\sigma}^d=F_{\sigma}^u \equiv {\cal{F}}_\sigma \normalsize \ee

\normalsize

\vspace{3mm}

\noindent From Eqs. (114)-(117) one gets the following useful
relationships to hold

\vspace{4mm}

\Large \be
\begin{array}{lcl} \frac{Y_{12}^q}{Y_{23}^q}
\frac{Y_{12}^d}{Y_{23}^d} \frac{{\cal{F}}_{\sigma}}{{\cal{F}}_{22}}
\thickapprox \frac{m_d\:m_b}{m_s^2} \qquad &, & \qquad
\frac{Y_{12}^q}{Y_{23}^q} \frac{Y_{12}^u}{Y_{23}^u}
\frac{{\cal{F}}_{\sigma}}{{\cal{F}}_{22}} \thickapprox \frac{m_u\:m_t}{m_c^2} \\
                                                             \\
\frac{Y_{12}^u}{Y_{12}^d} \thickapprox \frac{m_s}{m_d}
\frac{m_u}{m_c} \qquad & , & \qquad \frac{Y_{23}^u}{Y_{23}^d}
\thickapprox \frac{m_b}{m_s} \:\frac{m_c}{m_t}          \\
                                           \\
\frac{s_{12}^d}{s_{23}^d}  \thickapprox \frac{s_{12}^u}{s_{23}^u}
\thickapprox
\frac{Y_{12}^q}{Y_{23}^q}\:\frac{{\cal{F}}_{23}}{{\cal{F}}_{22}}
\qquad & , & \qquad \frac{s_{12}^u}{s_{12}^d} \thickapprox
\frac{s_{23}^u}{s_{23}^d} \thickapprox \frac{Y_{33}^u}{Y_{33}^d}\:.
\end{array}
 \normalsize \ee

\normalsize

\vspace{3mm}

\noindent The combination of Eqs. (114)-(118) yields

\large \be s_{13}^u  \thickapprox \left( \frac{Y_{33}^u}{Y_{33}^d}
\right)^2 s_{13}^d  + \left[ (\frac{m_c}{m_t})^2 - \left(
\frac{Y_{33}^u}{Y_{33}^d} \right)^2 (\frac{m_s}{m_b})^2 \right]\:
\frac{s_{12}^d}{s_{23}^d} \:.\normalsize \ee

\normalsize
\vspace{3mm}

\noindent Imposing now for the sake of simplicity

\be \frac{Y_{33}^u}{Y_{33}^d}=\frac{Y_{23}^u}{Y_{23}^d} \thickapprox
\frac{m_b}{m_s} \:\frac{m_c}{m_t} \equiv r \:,\ee

\vspace{3mm}

\noindent one reaches the simplified relationships between mixing
angles in the u and d quark sectors:

\be s_{12}^u \thickapprox r \:s_{12}^d \qquad , \qquad s_{23}^u
\thickapprox r \:s_{23}^d \qquad , \qquad s_{13}^u \thickapprox r^2
\:s_{13}^d \ee

\vspace{4mm}

Equation (121) allows one to write the $V_{CKM}=(V_L^u)^T \:V_L^d$
quark mixing matrix in terms of four parameters: $r$ and the three
mixing angles $s_{12}^d$, $s_{23}^d$, and $s_{13}^d$. A numerical
analysis shows that setting for instance $r = .317239712$,
corresponding to using the central values for the quark masses
$m_s$, $m_b$, $m_c$, and $m_t$ reported in the Particle Data
Group, Ref.\cite{pdg}, and the values $s_{12}^d = 0.32721$,
$s_{23}^d = 0.0604208$ and $s_{13}^d = 0.00921978$ yields the
quark mixing matrix (ignoring CP violation):

\vspace{4mm}
 \be V_{CKM}\approx \left( \begin{array}{rcr} 0.973776 & 0.227474 &
- 0.003963\\
- 0.227438 & 0.972890 & - 0.041912 \\
- 0.005678 & 0.041715 & 0.999113
\end{array} \right) \:,\ee

\vspace{3mm} \noindent Notice that except the matrix element
$V_{td}$, the other eight entries lie within the best fit range
values reported in Ref.\cite{pdg}. These results suggest that the
approach given in Eq.(64) for the orthogonal mixing matrices of
charged fermions is a good approximation.

\vspace{5mm}

\subsection{Quark-Lepton complementarity relations}

Using the quark mixing angles of Eq.(122) and the range of lepton
mixing angles of Eqs. (106)-(111) allows one to obtain the
following rough estimation for the quark-lepton complementary
relations\cite{qlcomrel}:

\vspace{3mm} \be \begin{array}{ccc}
\theta_{12}^{PMNS}+\theta_{12}^{CKM}& \thickapprox &
41.543^{\circ}\:-\:42.152^{\circ} \quad , \quad
41.543^{\circ}\:-\:43.139^{\circ} \quad , \quad 41.543^{\circ}\:-\:44.066^{\circ} \:,\\
                                                       \\
\theta_{23}^{PMNS}+\theta_{23}^{CKM}& \thickapprox &
36.835^{\circ}\:-\:37.184^{\circ} \quad , \quad
36.835^{\circ}\:-\:37.754^{\circ} \quad , \quad
36.835^{\circ}\:-\:38.295^{\circ} \:,
\end{array}  \ee

\vspace{3mm} \noindent for $s_{13}^2=0.034\:,\:0.037
\:,\text{and}\:0.04$, respectively.

\pagebreak

\section{FCNCs and rare decays for charged leptons}

The new exotic scalar particles introduced to implement the
radiative mass generation mechanism have the capability to induce
FCNCs and contribute to "flavor violation" processes such as $F
\rightarrow f_1 f_2 f_3$, to "radiative flavor violating"
processes such as $\mu \rightarrow e \gamma$, $\tau \rightarrow
\mu \gamma$, $\tau \rightarrow e \gamma$, as well as to the
"anomalous magnetic moments" (AMMs) of fermions. In this section I
compute roughly these additional contributions for the charged
leptons.

\vspace{3mm}

Once the generation of fermion masses is completed, the
transformations between gauge ($0$ superscript) and mass
(physical) eigenstates are for scalars $\Phi_i = U_{ij} \sigma_j$,
Eq.(17), for charged leptons

\be {\psi^0}_{eL,eR} = V_{eL,eR} \;{\psi}_{eL,eR}  \:, \ee

\noindent where

\be V_{eL,eR}= V_{eL,eR}^{(1)} V_{eL,eR}^{(2)} \qquad , \qquad
{\psi^0}_{eL,eR}^T = (e^0, \mu^0, \tau^0)_{L,R} \qquad , \qquad
\psi_{eL,eR}^T = (e, \mu, \tau)_{L,R} \:,\ee

\vspace{3mm} \noindent and analogous transformations for quarks.

\subsection{Lepton flavor violation (LFV) processes $F \rightarrow f_1 f_2 f_3$ }

The scalar fields $(\phi_9, \phi_{10}, \phi_{12}, \phi_{11})$
allow tree level flavor changing vertices through the couplings in
Eq.(7). In particular they may induce tree level "lepton flavor
violation" (LFV) processes such as  $\tau \rightarrow \mu \mu
\mu$, $\tau \rightarrow \mu \mu e$,  $\tau \rightarrow \mu e e$,
$\tau \rightarrow e e e$, and  $\mu \rightarrow e e e$. The
generic diagram for these processes is shown in Fig. 4. The decay
rate contribution from this generic diagram may be taken as
\cite{lfvdecays}

\be \Gamma(F \rightarrow f_1 f_2 f_3) \approx \frac{m_F^5
Y_l^4}{3072\:\pi^3 M_{\phi}^4} \;,\ee

\vspace{3mm} \noindent with $Y_l$ being a coupling constant.

\vspace{3mm}

\subsubsection{$\mu \rightarrow e e e$}

Here I discuss some details about the decay $\mu \rightarrow e e
e$. This rare decay is of particular interest to be analyzed
because experimentally it is strongly suppressed. The dominant
contribution to this decay comes from the diagrams of Fig. 5.
Then, from Eqs.(7) and (126), a rough estimation for this decay
rate may be written as

\be \Gamma(\mu \rightarrow e e e) \approx \frac{m_{\mu}^5
Y_{12}^4}{3072\:\pi^3} \frac{1}{2} \left\{ (V_{eL})_{21}^2 \left(
\sum_k \frac{U_{2k}}{M_k^2} \right)^2 + (V_{eR})_{21}^2 \left(
\sum_k \frac{U_{3k}}{M_k^2} \right)^2 \right\}_{\mu eee} \;, \ee

\vspace{3mm} \noindent and therefore the branching ratio for this
process is\footnote{I write the experimental bounds reported in
Particle Data Group Ref.\cite{pdg}.}

\be BR (\mu \rightarrow eee) \equiv \frac{\Gamma(\mu \rightarrow e
e e)}{(\Gamma_\mu)_T} \approx \frac{M_W^4}{g_w^4} Y_{12}^4 \{ \;\;
\}_{\mu eee} < 1 \times 10^{-12} \;,\ee

\vspace{3mm}

\noindent where I take

\be (\Gamma_\mu)_T \approx \Gamma(\mu \rightarrow \nu_\mu  e
\bar{\nu_e}) = \left( \frac{g_W m_\mu}{M_W} \right)^4
\frac{m_\mu}{12 (8 \pi)^3} \ee

\vspace{5mm}

\subsubsection{$\tau \rightarrow \mu \mu \mu$}

\be  \Gamma(\tau \rightarrow \mu \mu \mu)
 \approx  \frac{m_{\tau}^5 Y_{23}^4}{3072
 \:\pi^3} \frac{1}{2}   \left\{
(V_{eL})_{32}^2  \left( \sum_k \frac{U_{1k}}{M_k^2} \right)_L^2 +
(V_{eR})_{32}^2 \left(  \sum_k \frac{U_{4k}}{M_k^2} \right)_R^2
\right\}_{\tau \mu \mu \mu} \ee

\vspace{3mm}

\noindent
 Using the mean life of $\tau = (290.6 \pm 1.0) \times 10^{-15} \;s$, and hence $(\Gamma_\tau)_T = \frac{1}{\tau}
\approx 2.2711631 \times 10^{-12}\;GeV$, one gets the branching
ratio

\vspace{3mm}

\be BR (\tau \rightarrow \mu \mu \mu) \approx C_\tau {m_\tau}^4
Y_{23}^4 \{ \;\; \}_{\tau \mu \mu \mu}< 1.9 \times 10^{-7} \;, \ee

\vspace{3mm} \noindent with $C_\tau \equiv 4.107105 \times 10^6$

\subsubsection{$\tau^- \rightarrow \mu^+ \mu^- e^-$}

\be \Gamma(\tau^- \rightarrow \mu^+ \mu^- e^-) \approx
\frac{m_{\tau}^5 (Y_{12} Y_{23})^2}{3072\:\pi^3} \frac{1}{2} \left\{
\left( \sum_k \frac{U_{1k} U_{2k}}{M_k^2} \right)^2 + \left( \sum_k
\frac{U_{3k} U_{4k}}{M_k^2} \right)^2 \right\}_{\tau \mu \mu e} \;,
\ee

\vspace{5mm} \noindent with the branching ratio

\be BR (\tau^- \rightarrow \mu^+ \mu^- e^- ) \approx C_\tau
{m_\tau}^4 (Y_{12} Y_{23})^2 \{ \;\; \}_{\tau \mu \mu e}< 2 \times
10^{-7} \ee

\vspace{3mm}

\subsubsection{$\tau^- \rightarrow \mu^+ e^- e^-$}

\be \Gamma(\tau^- \rightarrow \mu^+ e^- e^-) \approx
\frac{m_{\tau}^5 (Y_{12} Y_{23})^2}{3072\:\pi^3} \frac{1}{2} \left\{
(V_{eL})_{21}^2 \left( \sum_k \frac{U_{1k} U_{2k}}{M_k^2} \right)^2
+ (V_{eR})_{21}^2 \left( \sum_k \frac{U_{3k} U_{4k}}{M_k^2}
\right)^2 \right\}_{\tau \mu e e} \;,\ee

\vspace{3mm}

\be BR (\tau^- \rightarrow \mu^+ e^- e^- ) \approx C_\tau
{m_\tau}^4 (Y_{12} Y_{23})^2 \{ \;\; \}_{\tau \mu e e}< 1.1 \times
10^{-7} \ee

\vspace{5mm}

\subsubsection{$\tau \rightarrow e e e$}

\be \Gamma(\tau \rightarrow e e e) \approx \frac{m_{\tau}^5 (Y_{12}
Y_{23})^2}{3072\:\pi^3} \frac{1}{2} \left\{ (V_{eL})_{21}^4 \left(
\sum_k \frac{U_{1k} U_{2k}}{M_k^2} \right)^2 + (V_{eR})_{21}^4
\left( \sum_k \frac{U_{3k} U_{4k}}{M_k^2} \right)^2 \right\}_{\tau
eee} \;,\ee

\vspace{3mm}

\be BR (\tau \rightarrow e e e ) \approx C_\tau {m_\tau}^4 (Y_{12}
Y_{23})^2 \{ \;\; \}_{\tau e e e}< 2 \times 10^{-7} \:.\ee

\subsection{Anomalous magnetic moments and radiative rare decays
$F \rightarrow f \gamma$}

The amplitude for the radiative process $f_1 \rightarrow f_2
\gamma$ with $f_1$ and $f_2$ being two equally charged fermions
and $\gamma$ a real photon is written as \cite{raddecays}

\be i {\cal{M}} ( f_1(p_1) \rightarrow f_2(p_2) + \gamma ) = i
{\bar{u}}_2(p_2) \left( \epsilon^{\mu} \gamma_{\mu} F_1^V(0)
\delta_{f_1f_2} + \frac{\sigma_{\mu\nu} q^{\nu}
\epsilon^{\mu}}{m_1+m_2} ( F_2^V(0) + F_2^A(0)\gamma_5 ) \right)
u_1(p_1) \:,\ee

\vspace{3mm} \noindent where $F_2^{V(A)}$ gives the AMM (electric
dipole moment) for the fermion $f_1$ when $f_1=f_2$. The generic
diagrams for the process $f_{1L} \rightarrow f_{2R} \;\gamma$ are
shown in Fig. 6, in these diagrams $\sigma$ stands for a mass
eigenstate scalar field. The respective evaluation of these
diagrams gives

\bn i A_L  \approx \frac{Y_l^2 q_e}{16 \pi^2} N(M_k ,m_i)
{\bar{e}}_{2R} i \sigma^{\mu\nu} q_{\nu} \epsilon_{\mu} e_{1L}
&\mbox{and}& i A_R \approx \frac{Y_l^2 q_e}{16 \pi^2} N(M_k ,m_i)
{\bar{e}}_{2L} i \sigma^{\mu\nu} q_{\nu} \epsilon_{\mu} e_{1R} \:,
\en

\vspace{3mm} \noindent where the second amplitude comes from the
diagrams where $L$ and $R$ are interchanged, and $N(M_k ,m_i)$ may
be approximated as

\be N(M_k ,m_i) \approx \frac{m_i}{M_k^2} \ln{\frac{M_k^2}{m_i^2}}
\ee

\vspace{3mm}

\noindent in the limit $M_k \gg m_i$. Notice that due to scalar
field mixing the contribution of these loops is finite as those in
the mass case.

\vspace{3mm}

Because of the fermion mixing matrices structure the diagrams that
make the largest contribution to the AMMs of the charged leptons
are, for the electron, the diagram with the muon inside the loop,
and for the muon and tau, the diagrams with tau as the internal
fermion.

\subsubsection{Muon anomalous magnetic moment}

The dominant  contribution for the muon AMM comes from the diagram
of Fig. 7, where the insertion of a photon on the internal lines
is understood as in the generic diagrams of Fig. 6. The expression
for this scalar contribution is \cite{raddecays, msher}

\vspace{3mm}

\Large \be \begin{array}{lll} a_\mu & = & \frac{m_\mu Y_{23}^2}{16
\pi^2} (V_{eL})_{22} (V_{eR})_{22} \left( G^{\mu_L} + G^{\mu_R}
\right) \approx \frac{m_\mu Y_{23}^2}{16 \pi^2} \left( G^{\mu_L} +
G^{\mu_R}  \right)                                 \\
                                                   \\
     & \approx & \frac{m_\mu m_\tau \:Y_{23}^2}{8 \pi^2} \sum_k \frac{U_{1k}
          U_{4k}}{M_k^2} \ln{\frac{M_k^2}{m_\tau^2}} \end{array}
          \;, \normalsize \ee

\normalsize \vspace{3mm}

\noindent where

\be \begin{array}{lll} G^{\mu_L}= G^{\mu_R} & = & \sum_{k,i} U_{1k}
U_{4k} (V_{eL})_{3i} (V_{eR})_{3i} N(M_k, m_i) \approx
\sum_{k} U_{1k}U_{4k} N(M_k, m_\tau)                 \\
                                                     \\
        & \approx & m_\tau \sum_k \frac{U_{1k}
          U_{4k}}{M_k^2} \ln{\frac{M_k^2}{m_\tau^2}} \;. \end{array} \ee

\subsubsection{Electron and tau anomalous magnetic moments}

\vspace{3mm}

Performing a similar analysis for $e$ and $\tau$ leptons, one gets

\vspace{3mm}

\noindent {\bf For electron:}

\bn a_e \approx \frac{m_e Y_{12}^2}{16 \pi^2} \left( G^{e_L} +
G^{e_R} \right) \approx \frac{m_e m_\mu \:Y_{12}^2}{8 \pi^2} \sum_k
\frac{U_{2k} U_{3k}}{M_k^2} \ln{\frac{M_k^2}{m_\mu^2}} \en

\vspace{3mm}

\noindent where

\be \begin{array}{lll} G^{e_L}= G^{e_R} & = & \sum_{k,i} U_{2k}
U_{3k} (V_{eL})_{2i} (V_{eR})_{2i} N(M_k, m_i) \approx \sum_{k}
U_{2k} U_{3k} N(M_k,m_\mu)                                 \\
                                                           \\
     & \approx & m_\mu \sum_k \frac{U_{2k}
          U_{3k}}{M_k^2} \ln{\frac{M_k^2}{m_\mu^2}} \;. \end{array} \ee

\vspace{5mm} \noindent {\bf For tau:}

\bn a_\tau \approx \frac{m_\tau Y_{33}^2}{16 \pi^2} \left(
G^{\tau_L} + G^{\tau_R} \right) \approx \frac{m_\tau^2 Y_{33}^2}{8
\pi^2} \sum_k \frac{U_{2k} U_{3k}}{M_k^2}
\ln{\frac{M_k^2}{m_\tau^2}} \en

\vspace{3mm}

\noindent where

\bn G^{\tau_L}= G^{\tau_R} & = & \sum_{k,i} U_{2k} U_{3k}
(V_{eL})_{3i} (V_{eR})_{3i} N(M_k, m_i) \approx \sum_{k} U_{2k}
U_{3k} N(M_k, m_\tau)             \nonumber\\
  &  & \approx m_\tau \sum_k \frac{U_{2k}
          U_{3k}}{M_k^2} \ln{\frac{M_k^2}{m_\tau^2}} \:.\en

\vspace{3mm}

\subsubsection{Radiative decay $\mu \rightarrow e \gamma $}

\vspace{3mm}

A similar analysis to the one for the muon AMM leads to the decay
rate

\be \begin{array}{lll} \Gamma(\mu \rightarrow e \gamma) & = & (m_\mu
+ m_e)^2 \frac{m_\mu (Y_{12} Y_{23})^2}{(16)^3\:\pi^5} (1 -
\frac{m_e}{m_\mu})^2  (1 - \frac{m_e^2}{m_\mu^2}) \left(
|(V_{eL})_{22} (V_{eR})_{11} G^{\mu_L e_R}|^2 + |(V_{eL})_{11}
(V_{eR})_{22} G^{\mu_R e_L}|^2 \right)                     \\
                                                            \\
 & \approx &  \frac{m_\mu^3 (Y_{12} Y_{23})^2}{(16)^3\:\pi^5}
\left( |G^{\mu_L e_R}|^2 + |G^{\mu_R e_L}|^2 \right)         \\
                                                             \\
    & \approx & \frac{m_\mu^3 m_\tau^2 (Y_{12} Y_{23})^2}{(16)^3\:\pi^5} \left\{
 (V_{eR})_{23}^2 \left( \sum_k \frac{U_{1k} U_{3k}}{M_k^2}
\ln{\frac{M_k^2}{m_\tau^2}} \right)^2 + (V_{eL})_{23}^2 \left(
\sum_k \frac{U_{2k} U_{4k}}{M_k^2} \ln{\frac{M_k^2}{m_\tau^2}}
\right)^2 \right\}_{\mu e \gamma} \;,\end{array} \ee

\vspace{3mm}

\be \begin{array}{c} G^{\mu_L e_R} =  \sum_{k,i} U_{1k} U_{3k}
(V_{eL})_{3i} (V_{eR})_{2i} N(M_k, m_i) \approx (V_{eR})_{23}
\:m_\tau \sum_k \frac{U_{1k} U_{3k}}{M_k^2}
\ln{\frac{M_k^2}{m_\tau^2}}\;,                  \\
                                                \\
 G^{\mu_R e_L} =  \sum_{k,i} U_{2k} U_{4k} (V_{eL})_{2i}
(V_{eR})_{3i} N(M_k, m_i) \approx (V_{eL})_{23} \:m_\tau \sum_k
\frac{U_{2k} U_{4k}}{M_k^2} \ln{\frac{M_k^2}{m_\tau^2}}\:.
\end{array} \ee

\vspace{3mm}

\noindent The resulting branching ratio may be expressed as

\be BR(\mu \rightarrow e \gamma) \; \approx \; \frac{3}{2 \pi^2}
\frac{m_\tau^2}{m_\mu^2} \left( \frac{M_W}{g_w}\right)^4 (Y_{12}
Y_{23})^2 \{ \;\; \}_{\mu e \gamma}
 \; < \; 1.2 \times 10^{- 11} \:.\ee

\subsubsection{Radiative decays $\tau \rightarrow \mu \gamma$ and $\tau \rightarrow e \gamma$}

\vspace{3mm} Carrying out a similar analysis, one gets

\vspace{3mm} {\bf  $\tau \rightarrow \mu \gamma$:}

\be \begin{array}{lll} \Gamma(\tau \rightarrow \mu \gamma) & \approx
& \frac{m_\tau^3 (Y_{23} Y_{33})^2}{(16)^3\:\pi^5} \left( |G^{\tau_L
\mu_R}|^2 + |G^{\tau_R \mu_L}|^2 \right)     \\
                                            \\
    & \approx & \frac{m_\tau^5 (Y_{23} Y_{33})^2}{(16)^3\:\pi^5}
\left\{ \left( \sum_k \frac{U_{2k} U_{4k}}{M_k^2}
\ln{\frac{M_k^2}{m_\tau^2}} \right)^2 +  \left( \sum_k \frac{U_{1k}
U_{3k}}{M_k^2} \ln{\frac{M_k^2}{m_\tau^2}} \right)^2 \right\}_{\tau
\mu \gamma} \end{array} \ee

 \vspace{3mm} \noindent where

\be \begin{array}{c} G^{\tau_L \mu_R}  =  \sum_{k,i} U_{2k} U_{4k}
(V_{eL})_{3i} (V_{eR})_{3i} N(M_k, m_i) \approx \:m_\tau \sum_k
\frac{U_{2k} U_{4k}}{M_k^2} \ln{\frac{M_k^2}{m_\tau^2}}\;, \\
                                                           \\
G^{\tau_R \mu_L}  =  \sum_{k,i} U_{1k} U_{3k} (V_{eL})_{3i}
(V_{eR})_{3i} N(M_k, m_i) \approx \:m_\tau \sum_k \frac{U_{1k}
U_{3k}}{M_k^2} \ln{\frac{M_k^2}{m_\tau^2}}\;, \end{array} \ee

\vspace{3mm} \noindent and branching ratio

\be BR(\tau \rightarrow \mu \gamma) \; \approx \; C_\tau^\prime
m_\tau^4 \:(Y_{23} Y_{33})^2 \{ \;\; \}_{\tau \mu \gamma}
 \; < \; 6.8 \times 10^{-8} \;,\ee

\vspace{3mm} \noindent with $C_\tau^\prime = 6.24205152 \times
10^5 \:.$

\vspace{10mm}

{\bf $\tau \rightarrow e \gamma$:}

\be \begin{array}{lll} \Gamma(\tau \rightarrow e \gamma) & \approx &
\frac{m_\tau^3 (Y_{12} Y_{23})^2}{(16)^3\:\pi^5} \left( |G^{\tau_L
e_R}|^2 + |G^{\tau_R e_L}|^2 \right)       \\
                                           \\
    & \approx & \frac{m_\tau^3 m_\mu^2 (Y_{12}
Y_{23})^2}{(16)^3\:\pi^5} \left\{ \left( \sum_k \frac{U_{1k}
U_{3k}}{M_k^2} \ln{\frac{M_k^2}{m_\mu^2}} \right)^2 +  \left(
\sum_k \frac{U_{2k} U_{4k}}{M_k^2} \ln{\frac{M_k^2}{m_\mu^2}}
\right)^2 \right\}_{\tau e \gamma} \end{array} \;,\ee

\vspace{3mm} \noindent where

\be \begin{array}{c} G^{\tau_L e_R} =  \sum_{k,i} U_{1k} U_{3k}
(V_{eL})_{2i} (V_{eR})_{2i} N(M_k, m_i) \approx \:m_\mu \sum_k
\frac{U_{1k} U_{3k}}{M_k^2} \ln{\frac{M_k^2}{m_\mu^2}}\;,    \\
                                                             \\
 G^{\tau_R e_L}  =  \sum_{k,i} U_{2k} U_{4k} (V_{eL})_{2i}
(V_{eR})_{2i} N(M_k, m_i) \approx \:m_\mu \sum_k \frac{U_{2k}
U_{4k}}{M_k^2} \ln{\frac{M_k^2}{m_\mu^2}}\;, \end{array} \ee

\vspace{3mm} \noindent and branching ratio

\be BR(\tau \rightarrow e \gamma) \; \approx \; C_\tau^\prime
m_\tau^2 m_\mu^2 \:(Y_{12} Y_{23})^2 \{ \;\; \}_{\tau e \gamma}
 \; < \; 1.1 \times 10^{-7} \;.\ee

\pagebreak

\section{Summary and conclusions}

I have reported a detailed analysis on fermion masses and mixing,
including neutrino mixing, within the context of an extension of
the standard model with an $U(1)_H$ flavor symmetry and
hierarchical radiative mass mechanism [2]. The results of this
analysis show that this model has the capability to accommodate
the observed spectrum of quark masses  and  mixing angles in the
$V_{CKM}$, as it is shown through the analysis in Secs. 3 and 6.
In a similar way the spectrum of charged lepton masses is
consistently generated through the analysis presented in Secs. 3
and 4. Upper bounds for the charged lepton mixing angles are given
in Eq. (83). These upper bounds imply that mixing in the lepton
sector comes almost completely from neutrino mixing; that is,
$V_{PMNS} \thickapprox V_\nu$. In this approach all lepton mixing
elements in $V_{PMNS}$ are written completely in terms of neutrino
masses. A numerical analysis shows that using $0.033 \lesssim
s_{13}^2 \lesssim 0.04$ one gets large mixing angles for
$\theta_{12}$ and $\theta_{23}$,
  $0.7 \leqslant \sin^2{2 \theta_{12}} \leqslant 0.777209$ and
 $0.87 \leqslant \sin^2{2 \theta_{23}} \leqslant 0.902305$, within
the present allowed $3 \sigma$ limits as reported by recent global
analysis of neutrino data oscillation
\cite{mohapatra,numixbggm,numixbgg}. Using these allowed ranges of
values for quark and lepton mixing, predictions for neutrino
masses and quark-lepton complementary relations are given in the
Eqs. (112), (113), and (123), respectively.

\vspace{3mm}

From the phenomenological point of view it is interesting to look
for a set of scalar mass parameters in $M_{\phi}^2$, Eq. (165),
that allows us to account for the strong experimental suppression
on LFV processes, such as $\mu \rightarrow e e e$, radiative rare
decays $\mu \rightarrow e \gamma$, $\tau \rightarrow \mu \gamma$,
$\tau \rightarrow e \gamma$, and the muon anomalous magnetic
moment. To achieve this goal a detailed numerical analysis and fit
it is needed, trying to keep at least the lowest scalar mass
eigenvalue $\eta_1$ within few TeV$^2$. However, it is important
to comment that Eq. (83) gives a good approximation for the upper
bounds on charged lepton mixing angles, and hence $V_{PMNS}
\thickapprox V_\nu$ would remain as a good approach in this model.

\vspace{3mm}

Thus, the contribution of my analysis in comparison to the one
realized in Ref. \cite{u1model} may be summarized in the following
aspects:

\begin{itemize}
\item {\it Scalar sector:}
      I have performed the analysis by considering the most general
      structure for $M_{\phi}^2$.
\item {\it Charged fermion sector:}
\begin{itemize}
\item I have obtained and then diagonalized the quark and lepton mass
      matrices at one and two loops in close analytical form.
\item Taking advantage of the strong hierarchy of quark and charged lepton masses,
      approximate expressions for the orthogonal mixing matrices of charged fermions are obtained.
\item I have reported general analytical expressions for the
branching ratios of LFV processes, radiative rare decays and for
the AMMs of charged leptons.
\end{itemize}
\item {\it Neutrinos:}
      The $V_{PMNS}$ lepton mixing matrix is obtained and
      written completely in terms of the neutrino masses, and
      numerical results for lepton mixing angles are provided.
\end{itemize}

\vspace{5mm}

{\bf Acknowledgments}

\vspace{5mm}

The author is thankful for support from the "Instituto
Polit\'ecnico Nacional" (Grants from EDI and COFAA) and the
Sistema Nacional de Investigadores (SNI) in Mexico.

\pagebreak

\section{Appendix: Diagonalization of a generic real symmetric 3x3 mass\\ matrix}

\vspace{5mm} \noindent In this appendix I give the details to
diagonalize a generic real symmetric mass matrix defined as

\be M \equiv \left(
\begin{array}{ccc}
    a & d & e \\ d & b & f \\ e & f & c \end{array} \right)\;.
\ee

\vspace{3mm}

\noindent One can diagonalize this matrix $M$ through the orthogonal
matrix $V$ as $V^T M V = diag(\lambda_1, \lambda_2, \lambda_3)$,
$\lambda_i$, $i=1,2,3$  being the eigenvalues of M. The determinant
equation $det|M-\lambda|=0$ imposes the constraint that each one of
the eigenvalues $\lambda_i$ obeys the cubic equation

\be - \lambda^3 + (a+b+c) \lambda^2 - (ab-d^2\;+ ac-e^2\;+bc-f^2)
\lambda \;+ abc - f^2 a - e^2 b - d^2 c + 2def = 0 \; .\ee

\vspace{3mm}

\noindent Thus, from Eq. (157) one obtains the following nonlinear
relationships to hold:

\be \begin{array}{rcl} \lambda_1 + \lambda_2 + \lambda_3 & = & a + b + c \quad \\
                                                     \\
\lambda_1\lambda_2 + \lambda_1\lambda_3 + \lambda_2\lambda_3 & = &
ab-d^2\;+ ac-e^2\;+ bc-f^2  \quad                    \\
                                                  \\
\lambda_1\lambda_2\lambda_3 & = & abc - f^2 a - e^2 b - d^2 c +
2def\:.
\end{array} \ee

\vspace{3mm} \noindent
 I do not impose any hierarchy between the eigenvalues
 $\lambda_1$, $\lambda_2$, and $\lambda_3$. However, I assume they are nondegenerated.
 Computing now the eigenvectors\footnote{Here still unnormalized},
the orthogonal matrix V may be writing as

\vspace{3mm}

\Large

\be V = \left( \begin{array}{ccc} x & y\:
\frac{F_1(\lambda_2)}{\Delta_2(\lambda_2)} & z\:
\frac{F_2(\lambda_3)}{\Delta_3(\lambda_3)} \\
     &              &                      \\
x\: \frac{F_1(\lambda_1)}{\Delta_1(\lambda_1)} & y & z\:
\frac{F_3(\lambda_3)}{\Delta_3(\lambda_3)} \\
     &              &                      \\
x\: \frac{F_2(\lambda_1)}{\Delta_1(\lambda_1)} & y\:
\frac{F_3(\lambda_2)}{\Delta_2(\lambda_2)} & z  \end{array} \right)
\:,    \normalsize        \ee

\vspace{3mm} \normalsize

\noindent where $x$, $y$, and $z$ are normalization constants, and
the functions involved  are defined as

\be \begin{array}{lcr}
 \Delta_1(\lambda)\equiv (b-\lambda)(c-\lambda) - f^2 & , &
 F_1(\lambda) \equiv - d (c-\lambda) + e f             \:,   \\
                                                         \\
\Delta_2(\lambda)\equiv (a-\lambda)(c-\lambda) - e^2  & , &
    F_2(\lambda) \equiv - e (b-\lambda) + d f          \;,   \\
                                                         \\
\Delta_3(\lambda)\equiv (a-\lambda)(b-\lambda) - d^2 & , &
    F_3(\lambda) \equiv - f (a-\lambda) + d e    \;.
\end{array} \ee

\vspace{3mm}

 Using properly Eqs. (157) and (158), it is possibly to check the
orthogonality between columns(eigenvectors) of $V$. Moreover, the
functions $\Delta_i(\lambda)$ and $F_i(\lambda)$ in Eq.(160) satisfy
the important and useful relationships

\vspace{3mm}

\be \begin{array}{ccc} F_1^2(\lambda) = \Delta_1(\lambda)
\Delta_2(\lambda)\;\;&,&\;\; F_1(\lambda)F_2(\lambda) =
\Delta_1(\lambda) F_3(\lambda)  \;, \\
                                \\
 F_2^2(\lambda) = \Delta_1(\lambda)
\Delta_3(\lambda)\;\;&,&\;\; F_1(\lambda)F_3(\lambda) =
\Delta_2(\lambda) F_2(\lambda)  \;, \\
                                \\
 F_3^2(\lambda) = \Delta_2(\lambda)
\Delta_3(\lambda)\;\;&,&\;\; F_2(\lambda)F_3(\lambda) =
\Delta_3(\lambda) F_1(\lambda) \;. \end{array} \ee

\vspace{2mm}

\noindent Defining

\be \begin{array}{lcl} h(\lambda) \equiv \Delta_1(\lambda) +
\Delta_2(\lambda) +\Delta_3(\lambda)&=& 3 \lambda^2 - 2
(a+b+c)\lambda + ab-d^2\;+ ac-e^2\;+ bc-f^2    \\
                                              \\
&=& 3 \lambda^2 - 2 ( \lambda_1 +\lambda_2 + \lambda_3) \lambda +
\lambda_1\lambda_2 + \lambda_1\lambda_3 + \lambda_2\lambda_3  \;,
\end{array} \ee

\vspace{2mm} \noindent explicitly,

\vspace{3mm}

\be \begin{array}{ccc} h(\lambda_1) &=& \lambda_1^2 - (\lambda_2 +
\lambda_3)\lambda_1 + \lambda_2\lambda_3 = (\lambda_2 -
\lambda_1)(\lambda_3 -
\lambda_1)                        \;,  \\
                                    \\
h(\lambda_2) &=& \lambda_2^2 - (\lambda_1 + \lambda_3)\lambda_2 +
\lambda_1\lambda_3 = (\lambda_1 - \lambda_2)(\lambda_3 -
\lambda_2)                        \;,  \\
                                    \\
h(\lambda_3) &=& \lambda_3^2 - (\lambda_1 + \lambda_2)\lambda_3 +
\lambda_1\lambda_2 = (\lambda_1 - \lambda_3)(\lambda_2 - \lambda_3)
\;. \end{array}  \ee

\vspace{5mm}

\noindent One can use now the relationships of Eqs. (161) and
(163) to normalize the eigenvectors, obtaining that in general the
square matrix elements $V_{ij}^2 \quad i,j=1,2,3$ may be expressed
as

\bn V_{ij}^2 = \frac{\Delta_i(\lambda_j)}{h(\lambda_j)} \geq 0 &
\mbox{and hence}& |V_{ij}| =
\sqrt{\frac{\Delta_i(\lambda_j)}{h(\lambda_j)}} \;.\en

\vspace{2mm}

Equation (164) defines the magnitudes for the matrix elements
$V_{ij}$ in Eq.(159). Setting now the diagonal elements of $V$ as
positives, $ x>0, y>0$, and $z>0$, the signs of the off diagonal
elements, $V_{ij},\: i \neq j$, may be obtained directly from the
Eq.(159) in a particular set of giving parameters $a, b, c, d, e$,
and $f$ that define the real symmetric mass matrix $M$ in Eq.
(156).

\vspace{3mm} It is important to mention here that the method
introduced in this Appendix to diagonalize a generic 3x3 real
symmetric mass matrix agrees with the diagonalization performed in
Ref. \cite{fritzsch} for the special case of Fritzsch's ansatz,
$a=e=0$.

\subsection{Diagonalization of the generic exotic scalar mass
matrices}

\vspace{3mm}

The most general square scalar mass matrix for the exotic scalar
fields, which mediate the radiative mass generation of the light
fermions at one and two loops in the u, d, and e charged fermion
sectors, may be written as

\be M_{\phi}^2 =\left( \begin{array}{cccc} a_1 & b
& 0 & 0\\ b & a_2 & c & 0\\ 0 & c & a_3 & d\\
0 &  0 & d & a_4
\end{array} \right).
\ee

\vspace{3mm}

\noindent This matrix may be diagonalized through the orthogonal
matrix U as $U^T M_\phi^2 U = diag(\eta_1, \eta_2, \eta_3,
\eta_4)$, $\eta_i \equiv M_i^2 \;, i = 1, 2, 3, 4$ being the
eigenvalues of $M_\phi^2$. Using the same procedure and method
introduced previously, the orthogonal matrix U may be writing as

\vspace{2mm}

\Large

\be U = \left( \begin{array}{cccc} x^{\prime} & y^{\prime}\:
\frac{f_2(\eta_2)}{\Delta_2(\eta_2)} & z^{\prime}\:
\frac{f_3(\eta_3)}{\Delta_3(\eta_3)} & t^{\prime}\:
\frac{f_4(\eta_4)}{\Delta_4(\eta_4)} \\
     &              &                      \\
x^{\prime}\: \frac{f_2(\eta_1)}{\Delta_1(\eta_1)} & y^{\prime} &
z^{\prime}\: \frac{g_3(\eta_3)}{\Delta_3(\eta_3)} & t^{\prime}\:
\frac{g_4(\eta_4)}{\Delta_4(\eta_4)}  \\
     &              &                      \\
x^{\prime}\: \frac{f_3(\eta_1)}{\Delta_1(\eta_1)} & y^{\prime}\:
\frac{g_3(\eta_2)}{\Delta_2(\eta_2)} & z^{\prime} & t^{\prime}\:
\frac{h_4(\eta_4)}{\Delta_4(\eta_4)}  \\
     &              &                 \\
x^{\prime}\: \frac{f_4(\eta_1)}{\Delta_1(\eta_1)} & y^{\prime}\:
\frac{g_4(\eta_2)}{\Delta_2(\eta_2)} & z^{\prime}\:
\frac{h_4(\eta_3)}{\Delta_3(\eta_3)} & t^{\prime}\:
\end{array} \right) \:, \normalsize \ee

\normalsize

\vspace{3mm}

\noindent where $x^{\prime}$, $y^{\prime}$, $z^{\prime}$, and
$t^{\prime}$ are normalization constants, and the functions
involved are defined as

\be \begin{array}{ccl} \Delta_1(\eta) & \equiv &
(a_2-\eta)(a_3-\eta)(a_4-\eta) - (a_2-\eta)d^2 - (a_4-\eta)c^2\;, \\
                                                      \\
 \Delta_2(\eta) & \equiv & (a_1-\eta)\left[(a_3-\eta)(a_4-\eta) - d^2 \right] \;,\\
                                                      \\
 \Delta_3(\eta) & \equiv & (a_4-\eta)\left[(a_1-\eta)(a_2-\eta) - b^2 \right] \;,\\
                                                      \\
\Delta_4(\eta) & \equiv & (a_1-\eta)(a_2-\eta)(a_3-\eta) -
(a_1-\eta)c^2 - (a_3-\eta)b^2  \;, \end{array} \ee

\vspace{3mm}

\noindent and

\vspace{3mm}

 \be \begin{array}{lcl} f_2(\eta)  \equiv  - b
\left[(a_3 - \eta)(a_4 - \eta) - d^2 \right]  & \quad , \quad &
g_3(\eta)  \equiv - c(a_1 - \eta)(a_4 - \eta) \;, \\
                                                  \\
f_3(\eta) \equiv  bc(a_4 - \eta)   & \quad , \quad &
 g_4(\eta)  \equiv  cd(a_1 - \eta)     \;,\\
                                                  \\
f_4(\eta)  \equiv  - bcd      & \quad , \quad &
 h_4(\eta)  \equiv  - d \left[(a_1 - \eta)(a_2 - \eta) - b^2
\right] \;. \end{array} \ee

\vspace{3mm} \noindent These functions satisfy relationships
analogous to those of Eq.(161), allowing us to obtain the
normalization constants and then to write the square matrix elements
as

\be U_{ij}^2 = \frac{\Delta_i(\eta_j)}{h(\eta_j)} \geqq 0
\qquad,\qquad \text{and hence} \qquad |U_{ij}|=
\sqrt{\frac{\Delta_i(\eta_j)}{h(\eta_j)}} \;, \ee

\noindent where

\be \begin{array}{ccc} h(\eta_1)  = (\eta_2 - \eta_1)(\eta_3 -
\eta_1)(\eta_4 - \eta_1)          & \quad , \quad &
 h(\eta_2)  =  (\eta_1 - \eta_2)(\eta_3 -
\eta_2)(\eta_4 - \eta_2)            \;, \\
                                      \\
 h(\eta_3)  =  (\eta_1 - \eta_3)(\eta_2 -
\eta_3)(\eta_4 - \eta_3)           & \quad , \quad &
 h(\eta_4)  =  (\eta_1 - \eta_4)(\eta_2 -
\eta_4)(\eta_3 - \eta_4) \;,
\end{array} \ee

\vspace{3mm} \noindent and leads to the useful equalities

\vspace{3mm}
 \be U_{1k} U_{4k}=\frac{f_4(\eta_k)}{h(\eta_k)} \quad , \quad  U_{2k} U_{4k}=\frac{g_4(\eta_k)}{h(\eta_k)} \quad ,
\quad U_{1k} U_{3k}=\frac{f_3(\eta_k)}{h(\eta_k)} \quad , \quad
U_{2k} U_{3k}=\frac{g_3(\eta_k)}{h(\eta_k)} \;.
 \ee

\pagebreak

\begin{figure}
\begin{center}
\includegraphics[width=0.9\columnwidth]{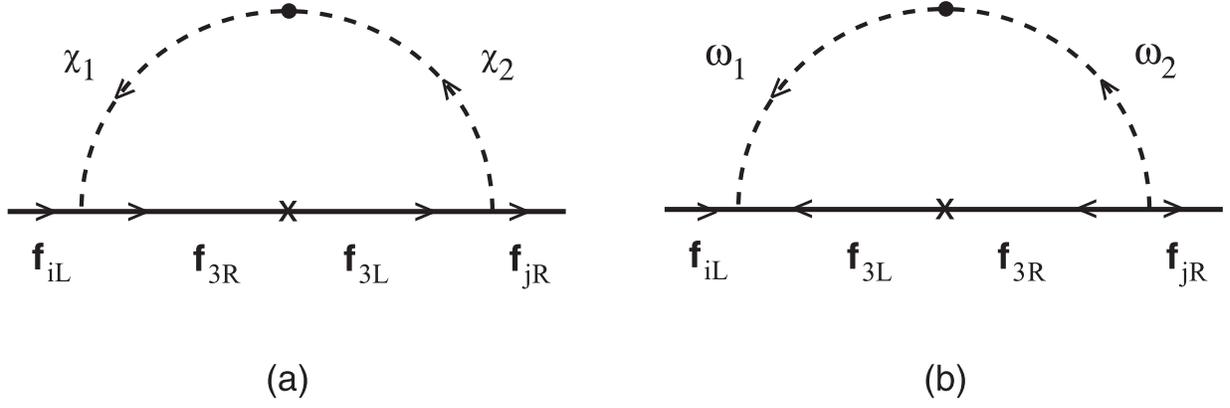}
\caption{Generic diagrams contributing to fermion masses. (a)
Dirac-type couplings, (b) Majorana-type couplings.}
 \end{center}
 \end{figure}

\begin{figure}
\begin{center}
\includegraphics[width=0.9\columnwidth]{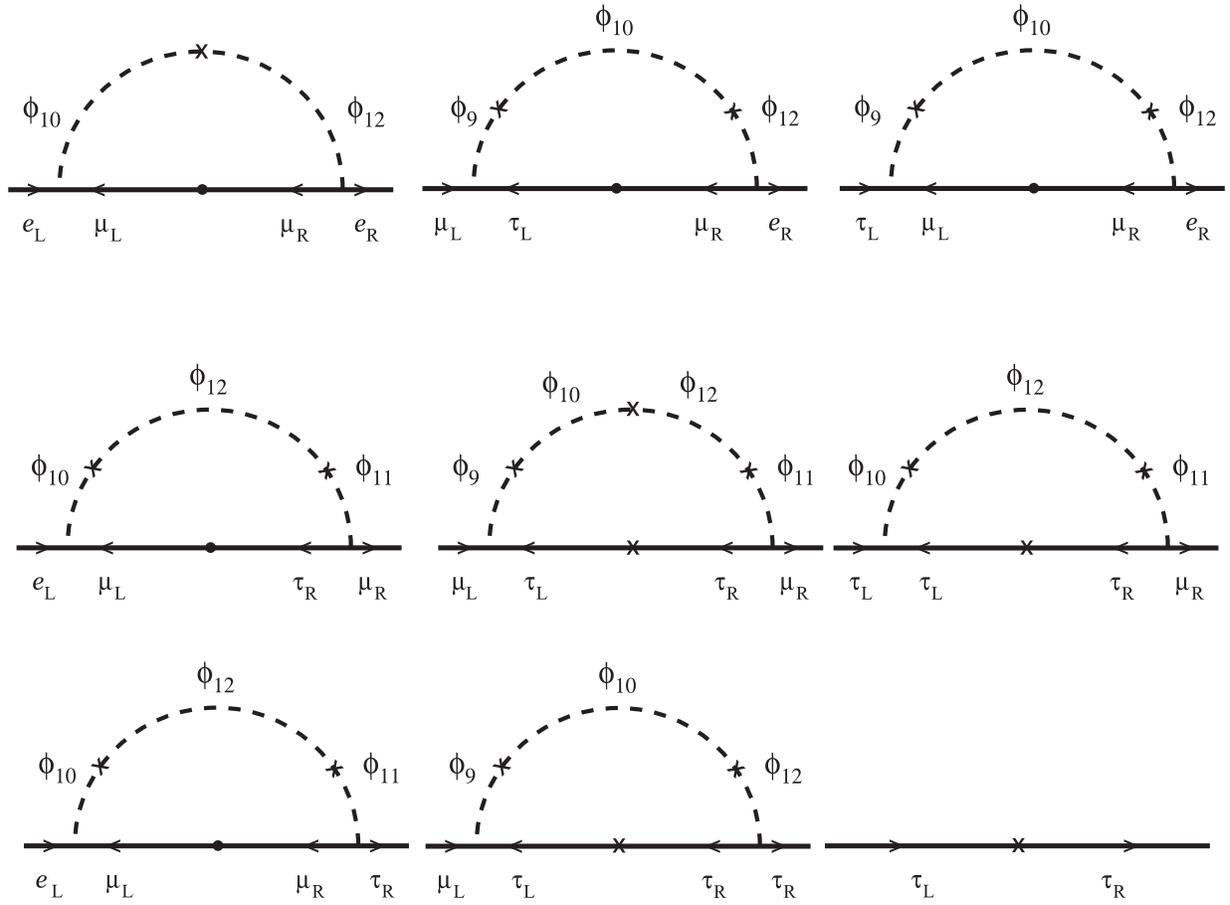}
\caption{Mass diagrams for the charged lepton sector.}
\end{center}
\end{figure}

\begin{figure}
\begin{center}
\includegraphics[width=0.9\columnwidth]{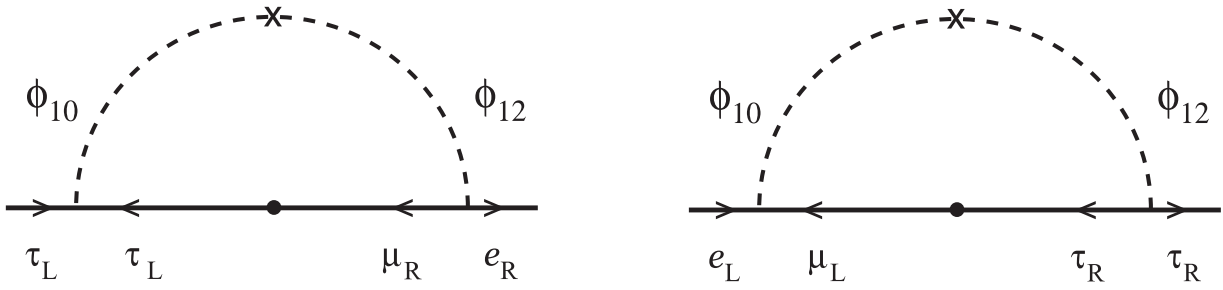}
\caption{Additional diagrams for the entries (1,3) and (3,1). }
\end{center}
\end{figure}

\begin{figure}
\begin{center}
\includegraphics[width=0.9\columnwidth]{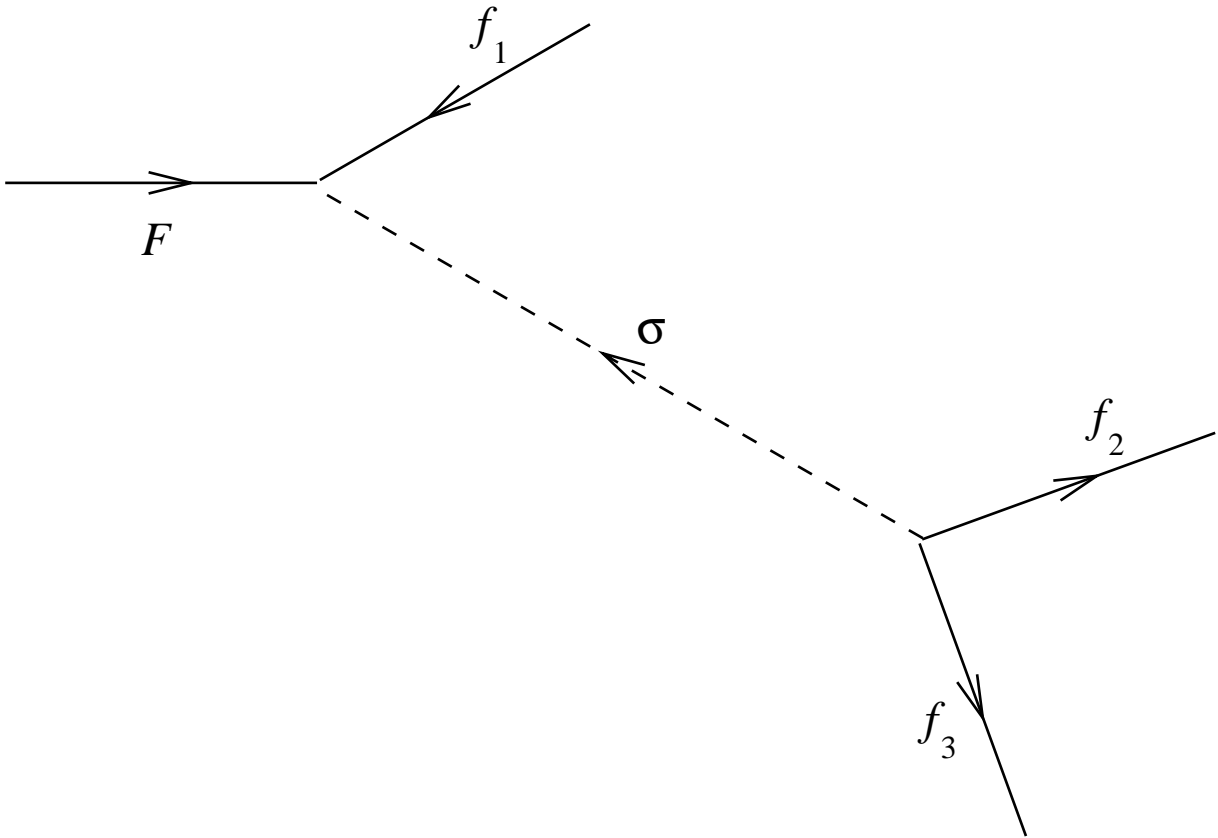}
\caption{Generic diagram for the LFV processes $F \rightarrow f_1
f_2 f_3$.}
\end{center}
\end{figure}

\begin{figure}
\begin{center}
\includegraphics[width=0.9\columnwidth]{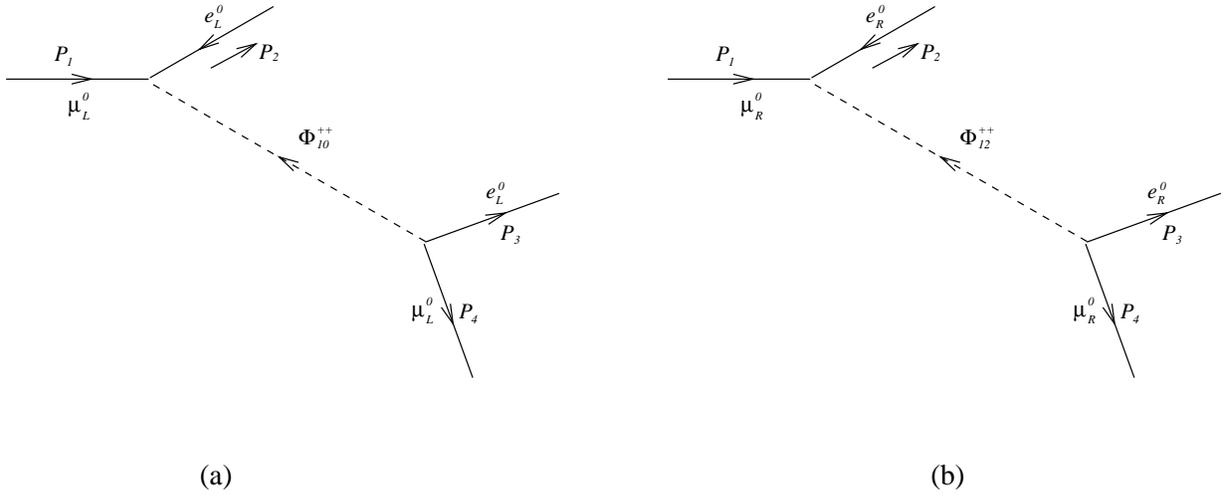}
\caption{Diagrams for the rare decay $\mu \rightarrow e e e$.}
\end{center}
\end{figure}

\begin{figure}
\begin{center}
\includegraphics[width=0.9\columnwidth]{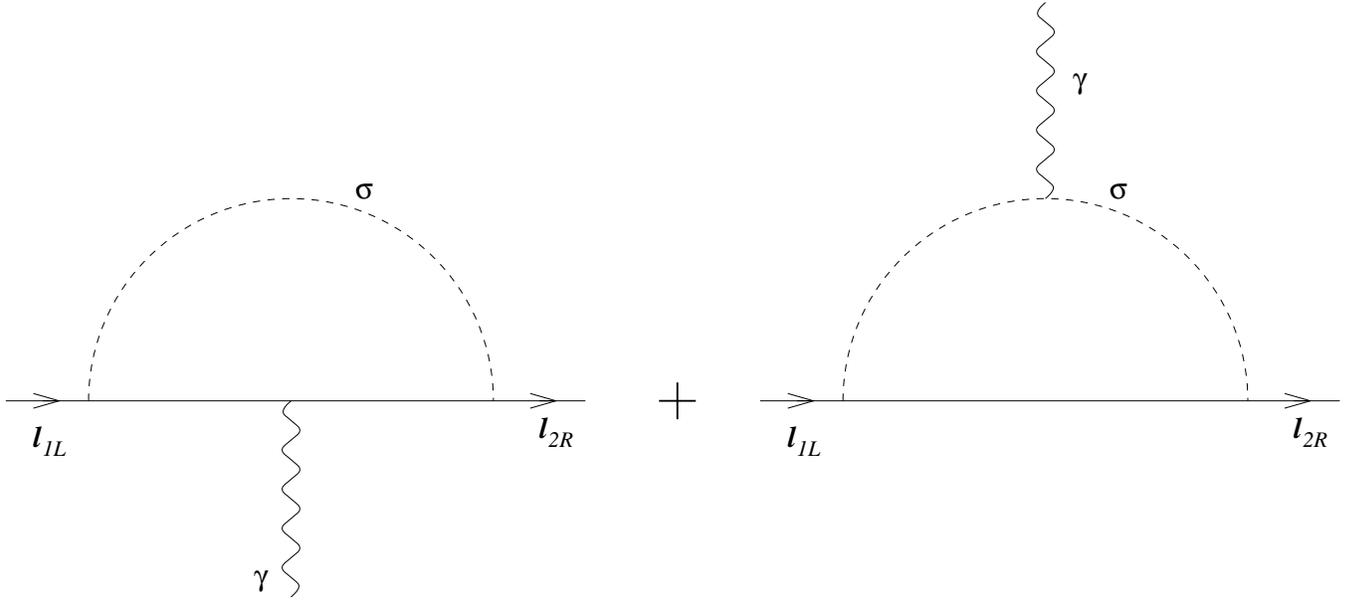}
\caption{Generic diagrams for the process $l_1 \rightarrow \l_2
\gamma$, where a scalar mass eigenstate $\sigma$ is involved.}
\end{center}
\end{figure}

\begin{figure}
\begin{center}
\includegraphics[width=0.9\columnwidth]{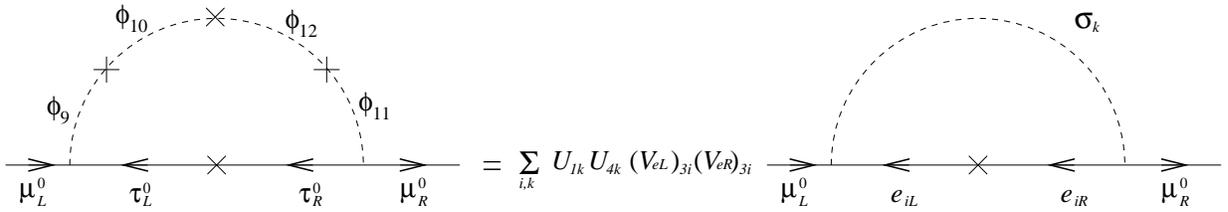}
\caption{Main contribution to the muon anomalous magnetic moment
(the insertion of a photon on the internal lines is understood as
in the Fig. 6).}
\end{center}
\end{figure}

\end{document}